\documentclass[a4paper,twocolumn,10pt,unpublished]{quantumarticle}
\pdfoutput=1
\usepackage[utf8]{inputenc}
\usepackage[english]{babel}
\usepackage[T1]{fontenc}
\usepackage[numbers,sort&compress]{natbib}

\usepackage{tikz}
\usepackage{lipsum}

\usepackage{graphicx}
\usepackage{floatpag}
\usepackage[toc]{appendix}
\usepackage{amssymb}
\usepackage{amsmath}
\usepackage{latexsym}
\usepackage{amsfonts}
\usepackage{mathtools}
\usepackage{braket} 
\usepackage[colorlinks,bookmarks=true,citecolor=green,linkcolor=green,urlcolor=blue]{hyperref}
\usepackage{amsthm}
\newcommand{\bs}[1]{\boldsymbol{\mathbf{#1}}}

\begin{document}

\title{Growing Schr\"odinger's cat states by local unitary time evolution of product states}

\author{Saverio Bocini}
\affiliation{Universit\'e Paris-Saclay, CNRS, LPTMS, 91405, Orsay, France}

\author{Maurizio Fagotti}
\email{maurizio.fagotti@universite-paris-saclay.fr}
\affiliation{Universit\'e Paris-Saclay, CNRS, LPTMS, 91405, Orsay, France}

\maketitle

\begin{abstract}
We envisage many-body systems that can be described by quantum spin-chain Hamiltonians with a trivial separable eigenstate. For generic Hamiltonians, such a state represents a quantum scar. We show that, typically, a macroscopically-entangled state naturally grows after a single projective measurement of just one spin in the quantum scar. Moreover, we identify a condition under which what is growing is a
``Schr\"odinger's cat state''. Our analysis does not reveal any particular requirement for the entangled state to develop, provided that the quantum scar does not minimise/maximise a local conservation law. 
We study two examples explicitly:  systems described by generic Hamiltonians and a model that exhibits a $U(1)$ hidden symmetry.
The latter can be reinterpreted as a 2-leg ladder in which the interactions along the legs are controlled by the local state on the other leg through transistor-like building blocks.   
\end{abstract}

Quantum superpositions of two macroscopically distinct states, known as ``Schr\"odinger's cat states'', from the famous Gedankenexperiment proposed by Schr\"odinger in 1935~\cite{Schrodinger1935Die,Trimmer1980The}, or also as ``(generalized) GHZ  states'', from the observation by Greenberger, Horne, and Zeilingerin~\cite{Greenberger1989Going} on quantum spin models escaping the original Bell's inequalities~\cite{Bell1964On}, are precious for their broad applications, e.g., in quantum metrology and quantum computation, but rare and generally short-lived for their fragility under real experimental conditions (which include decoherence, noise, particle loss, etc.). Several protocols have been proposed to contain the latter problems~\cite{AgarwalCat1997,Lombardo2015Deterministic,Wang2016A,Hacker2019Deterministic,Alexander2020Generating,Zhao2021Creation,Cosacchi2021Schrodinger,Wang2022A,Roscilde2022} and some of them have been experimentally realised~\cite{Gao2010Experimental,EtesseExperimental2015,Omran2019Experimental,SongExperimental}. While such an instability undermines   applications, it is, in fact, the  defining feature of  cat states, or, more generally, of states with extensive multipartite entanglement~\cite{Shimizu2002}. 
These are somehow unnatural states of matter in which fundamental physical properties such as cluster decomposition are lost. Engineering a cat state requires therefore a lot of control of the system, which, in turn, results in fine-tuned protocols designed with the clear goal of generating such exceptional states. 

We propose a theoretical protocol that stands out for its naturalness, in the sense that the growth of macroscopic entanglement is a manifestation of an intriguing physical phenomenon rather than of clever manipulations of a system. 
The most general system that we consider is described by a quantum spin-$\frac{1}{2}$ chain Hamiltonian $\bs H$ with local interactions and a trivial separable eigenstate.
Without loss of generality, the latter can be set equal to 
\begin{equation}
\ket{\Psi(0)}=\ket{\Uparrow}\equiv \ket{\uparrow\cdots\uparrow}\, ,
\end{equation}
where $\ket{\uparrow}$ denotes the eigenvector of the Pauli matrix $\sigma^z$ with eigenvalue~$1$ and $\Uparrow$ will represent in this paper a generic number of adjacent $\uparrow$. Often the presence of such a trivial eigenstate is a consequence of a $U(1)$ symmetry, e.g., the conservation of $\bs S^z=\frac{1}{2}\sum_\ell\bs\sigma_\ell^z$, where $\bs\sigma_\ell^\alpha$ are local operators acting like the Pauli matrices $\sigma^\alpha$ on site $\ell$ and like the identity elsewhere. As we will see before long, the most interesting cases are however those in which $\bs S^z$ is not conserved. If there are no additional relevant conserved operators, either $\ket{\Uparrow}$ is  the ground state or it is an exact \emph{quantum scar}~\cite{Bernien2017Probing,Turner2018Weak,Regnault2022Quantum}: its properties contrast with those of the eigenstates with similar energy. Quantum scars have recently attracted a lot of attention~\cite{moudgalya2021,Dooley2021Robust,scars_naturephys2022}, also in connection with their unusual entanglement properties,  displaying low bipartite entanglement, still potentially featuring extensive multipartite entanglement~\cite{Desaules2022Extensive}. In our case the quantum scar is exceptional, as it is fully separable. 

\begin{figure}
\centering
\includegraphics[width=0.9\linewidth]{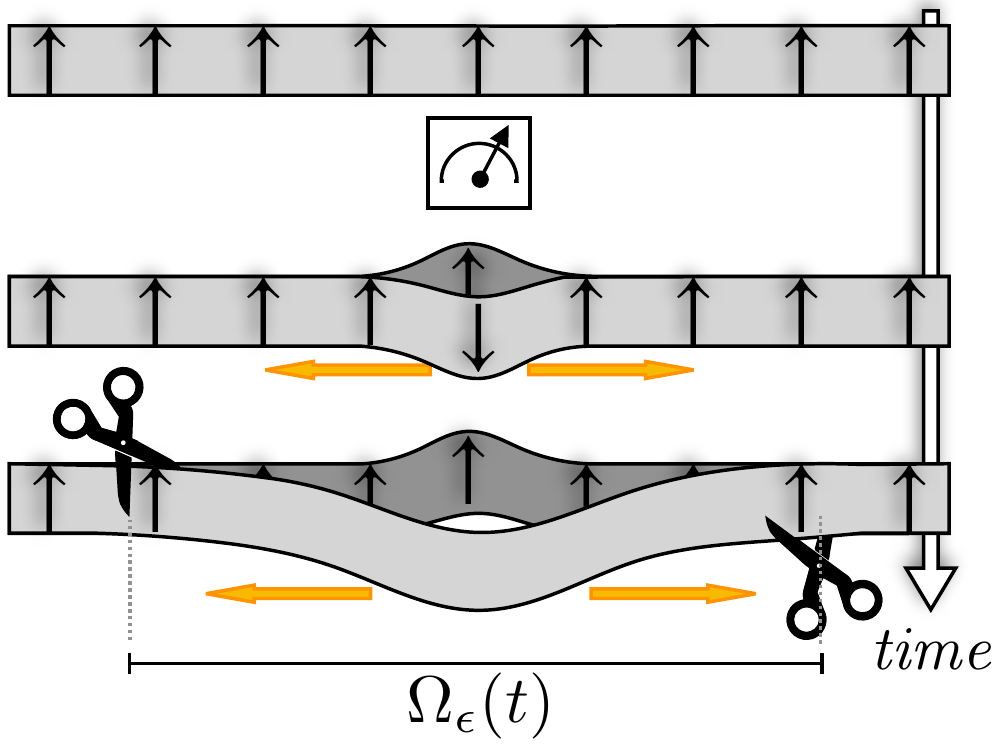}
\caption{Cartoon of the protocol. The state with a single spin down time evolves into a state that is macroscopically different from the pre-measurement state. If its fluctuations remain ordinary, the resulting superposition is a Schr\"odinger's cat state. The scissors represent the fact that we ignore the region outside $\Omega_\epsilon(t)$.
}
\label{fig:cat}
\end{figure}

The basic idea is that the quantum scar is essentially metastable and could be transmuted into a state with extensive multipartite entanglement just by a local perturbation. 
Specifically, we consider the effect of a quantum measurement of a spin in a tilted direction with respect to $z$, so that the state is projected into
\begin{equation}\label{eq:Psi0}
\ket{\Psi_{\theta}(0^+)}=e^{i\theta \bs \sigma_0^y}\ket{\Uparrow}=\cos\theta\ket{\Uparrow}+\sin\theta \ket{\Uparrow\downarrow \Uparrow}
\end{equation} 
with probability $\cos^2\theta$.
Time evolution affects only the second term on the right hand side of Eq.~[\ref{eq:Psi0}], hence
\begin{equation}\label{eq:Psi01}
\ket{\Psi_{\theta}(t)}=\cos\theta\ket{\Uparrow}+\sin\theta e^{-i \bs H t}\ket{\Uparrow\downarrow \Uparrow},\quad t>0\, .
\end{equation}
The Lieb-Robinson bounds~\cite{Lieb1972The} ensure that the perturbation is irrelevant outside a light cone emerging from the space-time point of the measurement. Since the state before the measurement is an eigenstate, the measurement triggers a nontrivial time evolution only within a region whose length grows linearly with the time. Hence, neglecting the part of the system outside that region does not significantly affect the purity of the state. 
We call $\Omega_\epsilon(t)$ the smallest subsystem that contains the spin that is measured and that, as clarified later, can be considered pure at time $t$ with accuracy $1-\epsilon$.
In the following, most of our considerations, including the use of terms such as ``macroscopic'' and ``extensive'', will be  referred to subsystem $\Omega_\epsilon(t)$. This allows us to treat on the same footing infinite systems, which are arguably more interesting from a theoretical point of view, and finite ones (provided that the actual system's size is larger than $\Omega_\epsilon(t)$), which are instead more relevant for the experiments.
The situation is pictorially described in Fig.~\ref{fig:cat}: after the spin flip at the origin, the component $\ket{\Uparrow\downarrow\Uparrow}$ of the state evolves in time, while the component $\ket{\Uparrow}$ stays constant; the effective size of the system is given by the spins that are inside the light cone originating from the local perturbation, while the rest of the spins are neglected.

First we show that, quite generally, in the settings considered the perturbation has everlasting effects on the expectation values of local observables arbitrarily far from the position of the measurement. That is to say, $e^{-i\bs H t}\ket{\Uparrow\downarrow\Uparrow}\equiv \ket{\Psi_{\pi/2}(t)}$ is macroscopically different from $\ket{\Uparrow}$, hence there exists an observable $\bs A=\sum_\ell \bs a_\ell$, where $\bs a_\ell$ are finite-support observables, such that $\braket{\Uparrow|\bs A|\Uparrow}-\braket{\Psi_{\pi/2}(t)|\bs A|\Psi_{\pi/2}(t)}$ is extensive, i.e., it scales linearly with the (effective) system size $|\Omega_\epsilon(t)|$ \cite{Morimae2005}.
Such an effect is already remarkable, as, generally,  local perturbations in quantum many-body systems either fade away or remain localised.  
We take it as the starting point of our investigation into the multipartite entanglement properties of $\ket{\Psi_\theta(t)}$.
We remind the reader that a state is said to have extensive multipartite entanglement if and only if the variance scales as the square of the system's length for at least one extensive operator~\cite{Hyllus2012Fisher,Toth2012,Frowis2018}.
Being $\ket{\Psi_{\pi/2}(t)}$ macroscopically different from $\ket{\Uparrow}$, the reader is already in the position to infer that the projective measurement on the quantum scar $\ket{\Uparrow}$ will generate a state with extensive multipartite entanglement. 
But the situation is even more interesting. 
We provide evidence and then conjecture that, if the model has a hidden $U(1)$ symmetry (see below for the definition), $\ket{\Psi_{\pi/2}(t)}$ is not macroscopically entangled; 
under those conditions, the projective measurement  generates a genuine Schr\"odinger cat's state consisting of an (ideally) arbitrarily large number of sites.

\section{The system}
We investigate a variety of Hamiltonians with local densities.
Our requirement of $\ket{\Uparrow}$ being eigenstate of the Hamiltonian allows for any kind of interactions that commute with the total spin in the $z$ direction, $\bs S^z = \frac{1}{2}\sum_\ell \bs\sigma^z_\ell$, (such as the hopping term $\bs\sigma_\ell^x\bs\sigma_{\ell+n}^x+\bs\sigma_\ell^y\bs\sigma_{\ell+n}^y$ or the longitudinal Dzyaloshinskii–Moriya interaction $\bs\sigma_\ell^x\bs\sigma_{\ell+n}^y-\bs\sigma_\ell^y\bs\sigma_{\ell+n}^x$) but not only that. The constraint is much weaker than a $U(1)$ symmetry and the Hamiltonian density can feature any term of the form $\frac{\bs 1-\bs\sigma_\ell^z}{2}\bs O_\ell\frac{\bs 1-\bs\sigma_\ell^z}{2}$, where $\bs O_\ell$ is a local operator with support in a region around $\ell$. Among the allowed operators,
those with a hidden $U(1)$ symmetry (such as $\bs\sigma_{\ell-1}^x(\bs 1-\bs\sigma_\ell^z)\bs\sigma_{\ell+1}^x$ or $\bs\sigma_{\ell-1}^x(\bs\sigma_\ell^x\bs\sigma_{\ell+1}^x+\bs\sigma_\ell^y\bs\sigma_{\ell+1}^y)\bs\sigma_{\ell+2}^x$) will play a special role.
\subsection{Hidden $U(1)$ symmetry}
The first class of Hamiltonians we consider has recently sparked some attention because it describes systems that are macroscopically sensitive to local perturbations~\cite{Fagotti2022Global,Fagotti2022Nonequilibrium}.
Such a sensitivity is triggered by so-called ``semilocal conservation laws'', whose densities act as local observables only in a restricted space of operators with a particular symmetry. 
This unusual property could seem, at first glance, innocuous, but it allows the state to retain memory of so-called ``string order''~\cite{derNijs89,Perez-Garcia2008String,Endres2011Observation}, making in turn symmetry-protected topological order survive the limit of infinite time~\cite{Fagotti2022Nonequilibrium}.
An example of operator with a  semilocal density in a system that is symmetric under the spin flip $\bs \sigma_\ell^{x,y}\rightarrow -\bs \sigma_\ell^{x,y}$ is
\begin{equation}\label{eq:tildeSz}
\tilde{\bs S}^z=\frac{1}{2}\sum_\ell\bs\Pi^z(\ell)\, ,
\end{equation}
where $\bs\Pi^z(\ell)$ can be thought of as a semi-infinite  string of $\bs\sigma^z$ and is such that $[\bs\Pi^z(\ell)]^2=\bs I$, $[\bs\Pi^z(\ell),\bs\sigma_j^z]=0$ and $\bs\Pi^z(\ell)\bs\sigma_j^{x,y}=\mathrm{sgn}(\ell-j-\frac{1}{2})\bs\sigma_j^{x,y} \bs\Pi^z(\ell)$. In spin-flip invariant settings, the density $\bs\Pi^z(\ell)$ of $\tilde{\bs S}^z$ is a local observable and, if $\tilde{\bs S}^z$ commutes with the Hamiltonian, it has important consequences on the evolution of the system, similarly to any other local conservation law --- we refer the reader to Ref.~\cite{Fagotti2022Nonequilibrium} for additional details. 

\begin{figure}
    \centering
\includegraphics[width=0.9\linewidth]{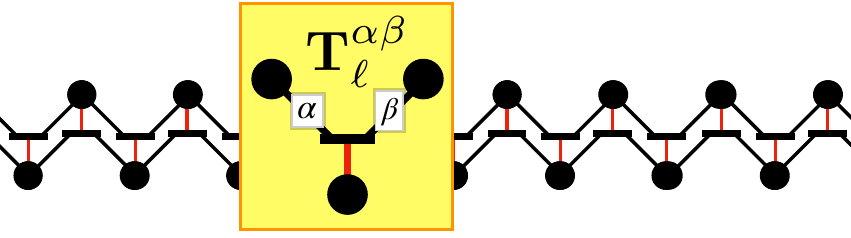}
    \caption{Pictorial representation of the quantum transistor chain, constructed as a sequence of building blocks $\bs T^{\alpha\beta}_\ell=
\bs\sigma_{\ell-1}^\alpha\frac{\bs I-\bs\sigma_\ell^z}{2}\bs\sigma_{\ell+1}^\beta$.}
    \label{fig:transistor_chain}
\end{figure}

The (two) symmetry sectors of the spin-flip symmetry are connected by any odd local operator, e.g., $\bs\sigma_j^x$. Acting with it on the state splits the latter in two parts where the expectation value of semilocal operators takes opposite signs. This is at the origin of the macroscopic effects of local perturbations pointed out in Ref.~\cite{Fagotti2022Global}. 
The behaviour of fluctuations  in these systems has not yet been investigated but the sensitivity of local observables to a single spin flip is a strong indication that the quantum measurement of a local observable such as $\bs\sigma_j^x$ could result in macroscopically entangled states.  
As a specific example, we consider the following Hamiltonian
\begin{multline}\label{eq:H1}
\bs H_1=\sum_\ell \frac{\bs 1-\bs \sigma_\ell^z}{8}[J\vec{\bs \sigma}_{\ell-1}\overset{S}{\cdot} \vec{\bs \sigma}_{\ell+1}
+\\+
\vec D\cdot (\vec{\bs \sigma}_{\ell-1}\times \vec{\bs \sigma}_{\ell+1})]
-\frac{\vec h}{2}\cdot\vec{\bs\sigma}_\ell
\end{multline}
where $\vec a\overset{S}{\cdot}\vec b=\vec a\cdot( S\vec b)$. This Hamiltonian commutes with the semilocal charge $\tilde{\bs S}^z$ --- Eq.~[\ref{eq:tildeSz}] --- provided that 
\begin{equation}\label{eq:H}
S=\begin{bmatrix}
\frac{1+\gamma}{2}&w&0\\
w&\frac{1-\gamma}{2}&0\\
0&0&\Delta
\end{bmatrix}\quad \vec D=\begin{bmatrix}
0\\
0\\
D^z
\end{bmatrix}\quad \vec h=\begin{bmatrix}
0\\
0\\
h^z
\end{bmatrix}\, .
\end{equation}
It describes a system that can be interpreted as a quantum transistor chain, in which the building blocks $\bs T^{\alpha\beta}_\ell=
\bs\sigma_{\ell-1}^\alpha\frac{\bs I-\bs\sigma_\ell^z}{2}\bs\sigma_{\ell+1}^\beta$, with $(\alpha,\beta)\in\{(x,x),(x,y),(y,x),(y,y),(z,z),(0,0)\}$ ($\bs\sigma^0_\ell\equiv \bs 1$), commute with  $\tilde{\bs S}^z$ ---  see Fig.~\ref{fig:transistor_chain}. 

It is known that $\bs H_1$ has infinitely many conserved operators with quasilocal~\cite{Ilievski2016Quasilocal} or semilocal~\cite{Fagotti2022Nonequilibrium} densities in the following regions of the parameter space:
\begin{enumerate}
\item \label{dualXXZ} If the only nonzero coupling constants are $J$, $\gamma$, and $h^z$, and $\gamma=\pm 1$, the model is dual to the (integrable) Heisenberg XXZ one~\cite{Zadnik2021The,Fagotti2022Global}; 
\item If the only nonzero coupling constants are $J$, $\Delta$, and $h^z$, the model belongs to the family of hard-rod deformations of XXZ studied in  Ref.~\cite{Pozsgay2021Integrable}. In particular, for $\Delta=0$ it is a special case of the Bariev model~\cite{Bariev1991Integrable} and was recently dubbed ``dual folded XXZ''~\cite{Zadnik2021The}. These are special integrable models exhibiting Hilbert-space fragmentation and quantum jamming~\cite{Zadnik2021The,Zadnik2021TheII,Bidzhiev2022Macroscopic,Zadnik2021Measurement}.
\end{enumerate}

For the first family of integrable models, Ref.~\cite{Fagotti2022Global} has already shown that a subsystem of $\ket{\Psi_{\pi/2}(t)}$, linearly growing  in time, is macroscopically different from the corresponding one in $\ket{\Uparrow}$. An analogous conclusion was drawn in Refs~\cite{Bidzhiev2022Macroscopic,Zadnik2021Measurement} for the second family of integrable models, though starting from slightly different product states (in those systems $\ket{\Uparrow\downarrow\Uparrow}$ is an excited eigenstate as well). Ref.~\cite{Bidzhiev2022Macroscopic} also observed that a similar behaviour could be seen even in generic systems with a jamming sector. 
We announce that a part of the phenomenology common to all these settings is much more general than the specific models studied so far.

\subsection{Generic model}
We establish contact with the systems currently studied in quantum simulators, for example with trapped ions~\cite{Monroe2021Programmable}, by considering the following Hamiltonian
\begin{equation}\label{eq:H2}
\bs H_2=\frac{1}{4}\sum_\ell\Bigl[\sum_{r=1}J_r\vec{\bs\sigma}_\ell\overset{S_r}{\cdot}\vec{\bs\sigma}_{\ell+r}+\vec D_r\cdot(\vec{\bs\sigma}_\ell\times \vec{\bs\sigma}_{\ell+r})\Bigr]-\frac{\vec h}{2}\cdot\vec{\bs\sigma}_\ell .
\end{equation}
This is the most general Hamiltonian where spins interact in pairs: there's a Heisenberg exchange term, a Dzyaloshinskii-Moriya interaction, and a coupling with an external field. 
The trivial quantum scar $\ket{\Uparrow}$ appears when
\begin{equation}
S_r=\begin{bmatrix}
1&0&\frac{1}{2}\gamma_r^{x}\\
0&1&\frac{1}{2}\gamma_r^{y}\\
\frac{1}{2}\gamma_r^{x}&\frac{1}{2}\gamma_r^{y}&1+\gamma_r^z
\end{bmatrix}\quad 
\vec h=\begin{bmatrix}
\frac{1}{2}\sum_{r=1} J_r \gamma^x_r\\
\frac{1}{2}\sum_{r=1} J_r \gamma^y_r\\
h^{z}
\end{bmatrix} .
\end{equation} 
The parameter $\vec \gamma$ incorporates both an anisotropy in the Heisenberg interaction and a rotation of the axes (which is relevant because we have fixed the orientation of the spins of the separable eigenstate).
As will be clarified before long, the effect we exploit to generate macroscopically entangled states requires a $U(1)$-breaking interaction, therefore we will only consider systems in which some coupling constants among $D_r^{x}$, $D_r^{y}$, $\gamma_r^x$, $\gamma_r^y$ are nonzero.

In such a generic setting there is no a-priori reason to expect the localised perturbation to produce macroscopic effects; for example, the system does not exhibit semilocal charges or special constraining interactions. Is the presence of a simple quantum scar sufficient to trigger the phenomenon?

Before reporting the results of our investigation, we remind the reader of some special regions of the parameter space:
\begin{itemize}
\item If the only nonzero coupling constants are $J_1$, $\gamma_1^z$, and $h^z$, the system is integrable and known as XXZ model, which is arguably the most important paradigm of quantum magnetism in 1D~\cite{Mikeska2004One}\footnote{Sometimes, especially in experimental works, the name ``XXZ model'' is referred also to the higher range model with nonzero $J_r$ and $\gamma_r^z=\gamma_1^z$.}. 
\item More generally, if $\vec D_r=0$ and $\vec \gamma_r=\vec\gamma_1$, the system describes an XYZ Heisenberg model with a tilted orientation, the integrable case corresponding to $\vec h=0$ and $J_r=\delta_{r 1 } J_1$. See also Ref.~\cite{Muller1985Implications} for a special region of the parameter space in which $\ket{\Uparrow}$ becomes the symmetry-breaking ground state.   
\item If the only nonzero coupling constants are $J_1$, $\gamma_1^x$, $D_1^y$, and $h^z$, in the limit $J_1\rightarrow 0$ at fixed $\gamma_1^x J_1=-2 D_1^y$ the system approaches the so-called quantum East model~\cite{Horssen2015Dynamics}, which has recently attracted a lot of attention for its unusual properties~\cite{Pancotti2020Quantum}. 
\end{itemize}

\section{Results}

\subsection{The effective system $\Omega(t)$}

We have introduced $\Omega_\epsilon(t)$ as the smallest subsystem, containing the measured spin, that can be considered pure at time $t$ with accuracy $1-\epsilon$. We clarify here this point.
First of all, it can be readily proven that the initial perturbation is irrelevant outside a light cone emerging from the space-time point of the measurement using a corollary of the Lieb-Robinson bounds derived in Ref.~\cite{Bravyi2006Lieb-Robinson}. Indeed, the Heisenberg representation of a local operator is exponentially close to an operator with support in a finite region including the support of the operator, the argument of the exponential being proportional to $d-v_{LR}|t|$, where $d$ is the smallest distance between the observable and the boundary of the region and $v_{LR}$ is the Lieb-Robinson velocity. In our specific case, this means that the following decomposition holds: $\exp[i\theta \bs \sigma_0^y(-t)]=\bs U_\Omega+\bs\Delta_\Omega$, where $\bs U_\Omega$ is a unitary operator with support in a region $\Omega$ centered at the measurement position and $\parallel\bs \Delta_\Omega\parallel\lesssim \exp[-(|\Omega|/2-v t)/\xi]$, with $\xi$ a non-universal constant. This gives
\begin{multline}
\parallel\ket{\Psi(t)}\bra{\Psi(t)}-\bs U_\Omega\ket{\Uparrow}\bra{\Uparrow}\bs U_\Omega^\dag\parallel=\\ \parallel\bs U_\Omega\ket{\Uparrow}\bra{\Uparrow}\bs\Delta_\Omega^\dag+\bs \Delta_\Omega\ket{\Uparrow}\bra{\Uparrow}\bs U_\Omega+
\bs \Delta_\Omega\ket{\Uparrow}\bra{\Uparrow}\bs\Delta_\Omega^\dag\parallel
\\\leq
2\parallel\bs \Delta_\Omega\parallel+\parallel\bs \Delta_\Omega\parallel^2\lesssim e^{-(|\Omega|/2-v t)}\, ;\nonumber
\end{multline}
that is to say, up to exponentially small corrections the reduced density matrix of $\Omega$ is pure.

We call $\Omega_\epsilon(t)$ the smallest spin block for which \mbox{$\parallel\mathrm{tr}_{\Omega_\epsilon(t)}[e^{-i \bs H t}\ket{\Uparrow\downarrow \Uparrow}\bra{\Uparrow\downarrow \Uparrow}e^{-i \bs H t}]-(\ket{\Uparrow}\bra{\Uparrow})_{\overline{\Omega_\epsilon(t)}}\parallel$} is smaller than $\epsilon$, where $(\ket{\Uparrow}\bra{\Uparrow})_{\overline{\Omega_\epsilon(t)}}$ is the state $\ket{\Uparrow}$ restricted to the complement of $\Omega_\epsilon(t)$. 
Note that $\Omega_\epsilon(0)$ contains only the spin at the origin for any choice of~$\epsilon$.

\subsection{$U(1)$ symmetry}
We point out here that, if the total magnetisation $\bs S^z$ is conserved, the states $\ket{\Uparrow}$ and $\ket{\Psi_{\pi/2}(t)}$ can not become macroscopically different and their linear combination is not macroscopically entangled. 
Specifically, Appendix~\ref{app:U(1)} provides a proof that a state $\ket{\Psi}$ obtained by time evolution under a $U(1)$-symmetric Hamiltonian after a local perturbation to $\ket{\Uparrow}$
is macroscopically equivalent to $\ket{\Uparrow}$; that is to say, the quantity $\braket{\Psi|\bs O|\Psi}-\braket{\Uparrow|\bs O|\Uparrow}$ is subextensive for any translational-invariant operator $\bs O = \sum_\ell \bs O_{A_\ell}$, with $\bs O_{A_\ell}$ local operators.
We also show that the variance of any of such operators $\bs O$ with respect to the state $\ket{\Psi}$ grows at most linearly with system size $|\Omega_\epsilon(t)|$, ruling out the possibility to obtain a macroscopically entangled state from time evolution of a locally-perturbed $\ket{\Uparrow}$. And this holds true for any local perturbation.
Physically, we can understand this result from the fact that, first, a local operator can only flip a finite number of spins, and, second, the initial state, $\ket{\Uparrow}$, has maximal $\bs S^z$. As a result, the Hilbert space accessible to time evolution is too small for macroscopic entanglement to develop.
This is the first clue supporting the expectation that, in order to build up macroscopic entanglement, $\ket{\Uparrow}$ 
can not be the ground state of a local conservation law. In generic systems, we can read this as a statement that $\ket{\Uparrow}$ should be a quantum scar.

\subsection{Hidden $U(1)$ symmetry}
We start with $\bs H_1$ --- Eq.~[\ref{eq:H1}] --- in the integrability region in which, as discussed before, we can take for granted  that $\ket{\Psi_{\pi/2}(t)}$ becomes macroscopically different from~$\ket{\Uparrow}$:
\begin{equation}
|\braket{\Uparrow\downarrow\Uparrow|e^{i \bs H_1 t}\bs O e^{-i \bs H_1 t}|\Uparrow\downarrow\Uparrow}-
\braket{\Uparrow|\bs O|\Uparrow}|\sim t\, ,
\end{equation}
with $\bs O$ a generic (spin-flip invariant) extensive operator. This is expected as long as the initial state  $\ket{\Uparrow}$ is in the middle of the spectrum of $\bs H_1$. Incidentally, we can understand that this latter condition is necessary by considering the dual XXZ model --- Eq.~[\ref{dualXXZ}] --- in which, for $|h^z|$ large enough, $\ket{\Uparrow}$ becomes the ground state (or the maximum energy state). By applying the Kramers-Wannier duality mapping proposed in Ref.~\cite{Fagotti2022Global} the system is mapped into the time evolution of a domain wall in the XXZ model, where the magnetic field $h^z$ plays the role of the anisotropy. As proven in Ref.~\cite{Mossel2010Relaxation}, the domain wall does not spread when the anisotropy is larger than a critical value, which is equivalent to say that the perturbation remains localised when $|h^z|$ is large enough to move $\ket{\Uparrow}$ at the boundaries of the energy spectrum.

In the integrability region of $\bs H_1$ there is just one piece of information that we cannot retrieve from the scientific literature:  the behaviour of fluctuations. This is important to us because the Lieb-Robinson bound guarantees $|\Omega_\epsilon(t)|=\mathcal O(t)$, and hence, if the variance of extensive observables  in $\ket{\Psi_{\pi/2}(t)}$ were $\mathcal O(t)$, we could readily conclude that the projective measurement generates a genuine cat state. 

\begin{figure}[!t]
\centering
\includegraphics[width=0.9\linewidth]{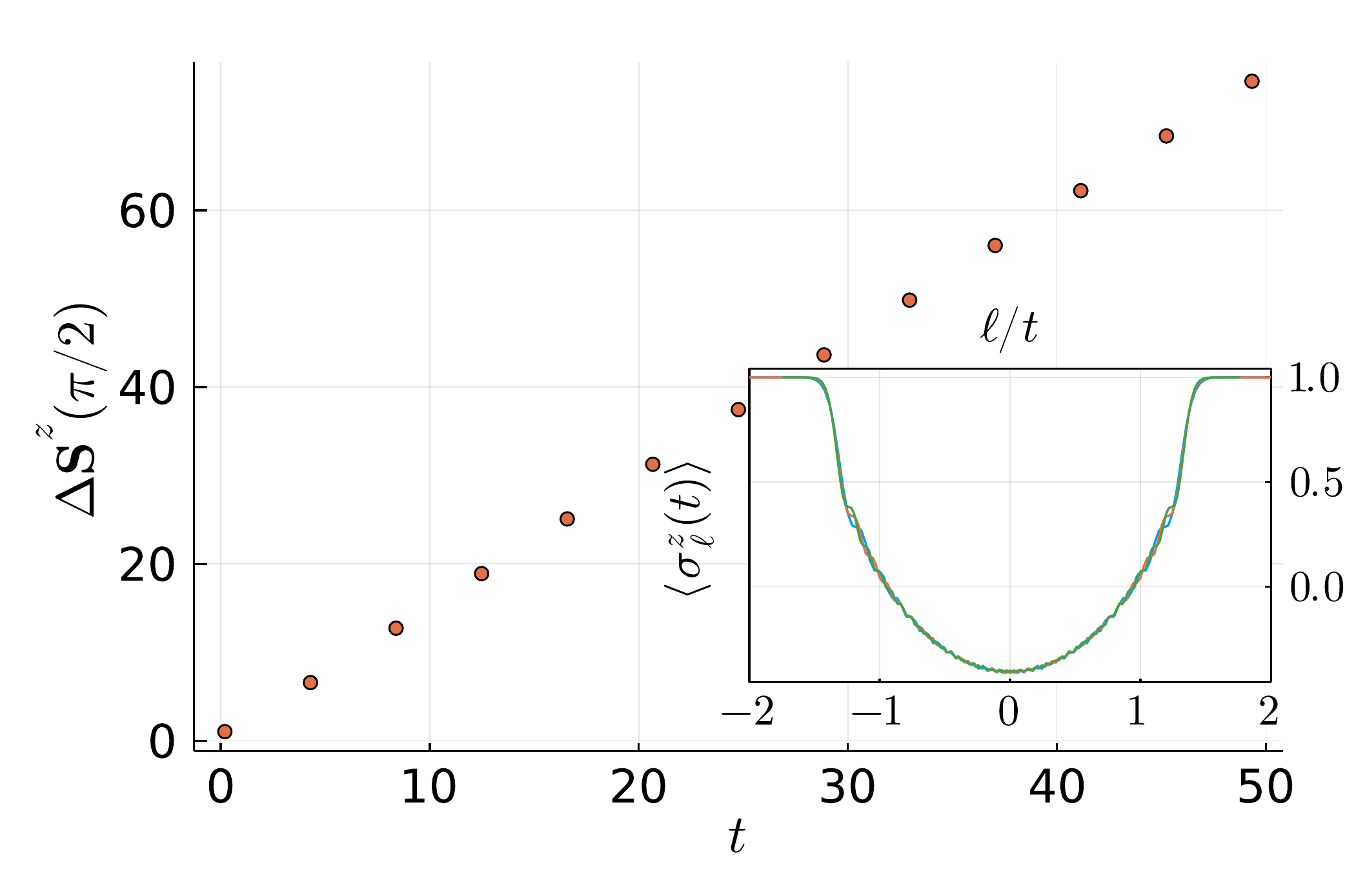}
\includegraphics[width=0.9\linewidth]{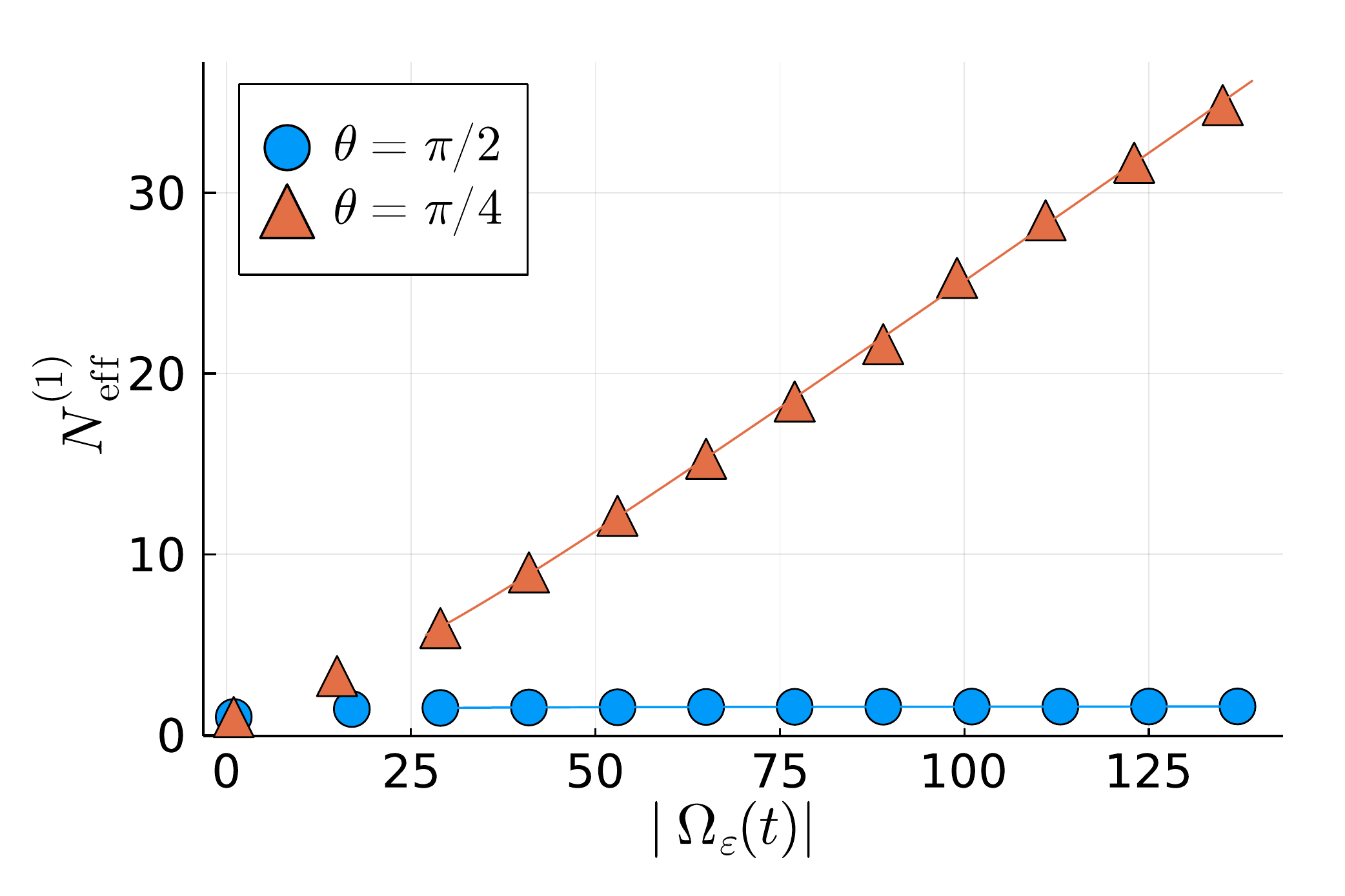}
\includegraphics[width=0.9\linewidth]{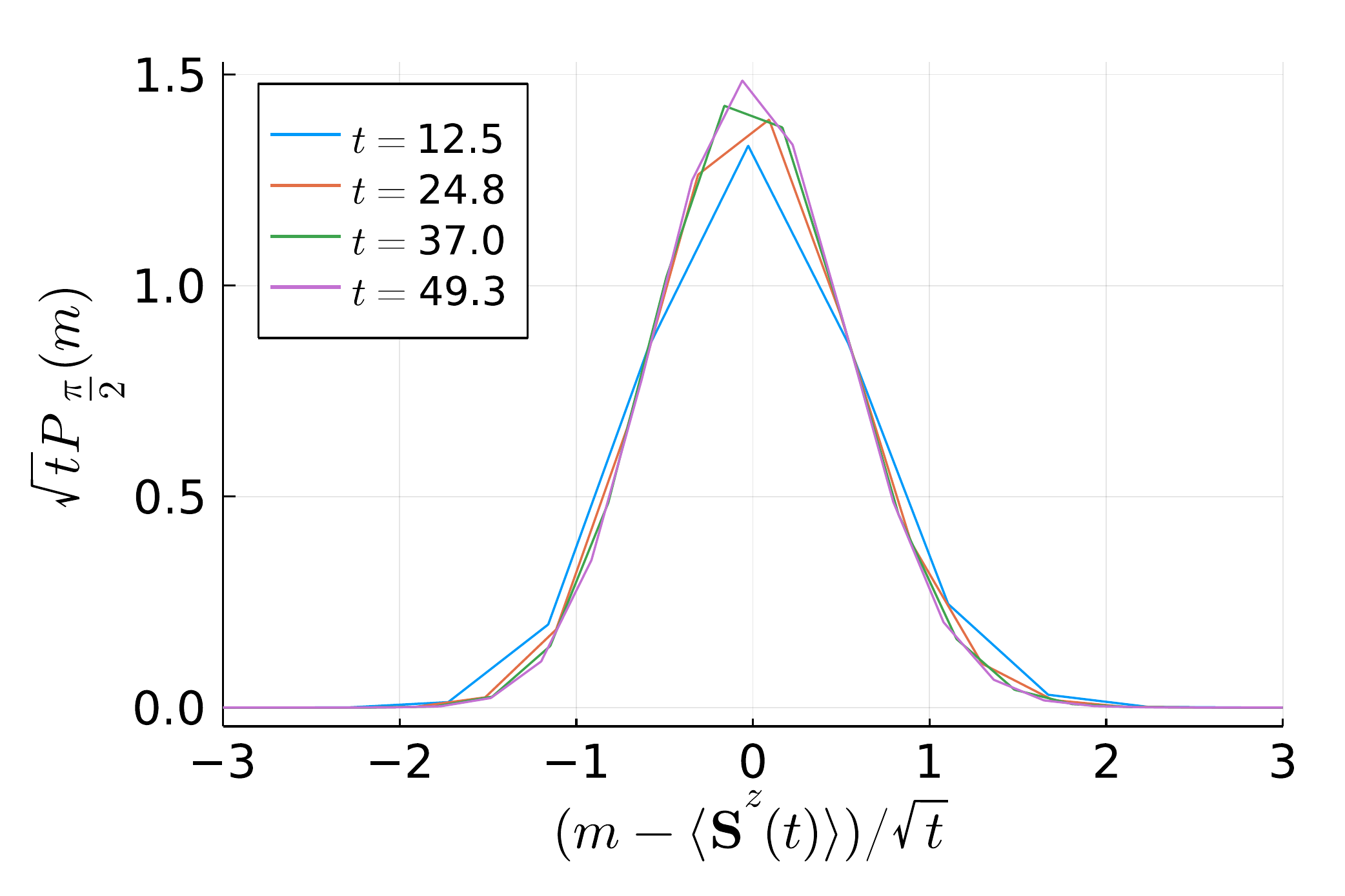}
\includegraphics[width=0.9\linewidth]{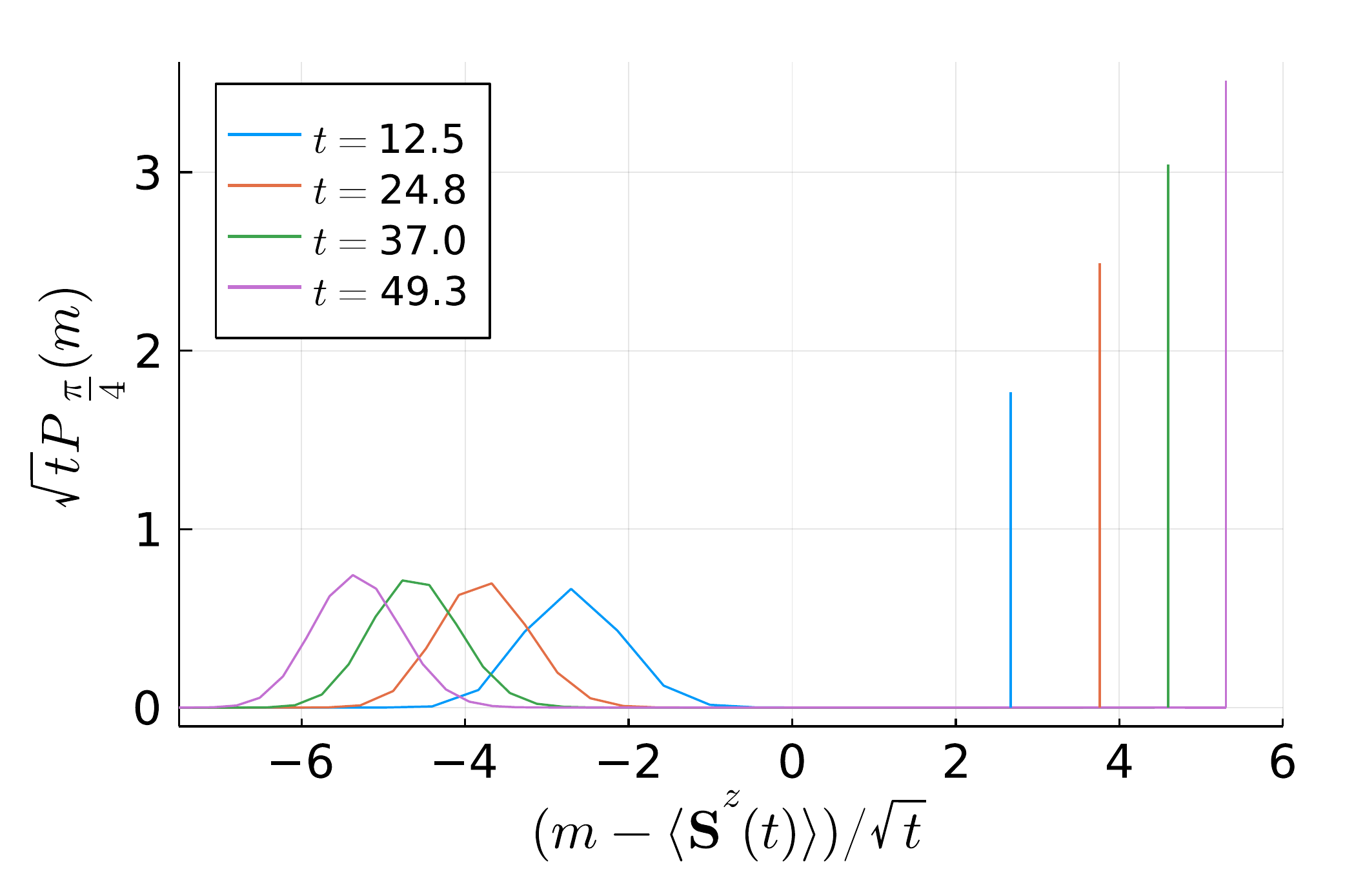}
\caption{Time evolution of $\ket{\Psi_\theta(t)}$ under Hamiltonian $\bs H_1$ with  $J=2.8$, $\gamma=1$, $w=\Delta=D^z=h^z=0$. 
From the top we have: 1) Difference between the total magnetization in $\ket{\Uparrow}$ and $\ket{\Psi_{\pi/2}(t)}$, i.e. $\Delta\bs S^z(\theta)=\braket{\Uparrow|\bs S^z|\Uparrow} - \braket{\Psi_\theta(t)|\bs S^z|\Psi_\theta(t)}$; inset:  local magnetization profile for the last four times of the main plot.
2) Macroscopic quantumness $N_{\mathrm{eff}}^{(1)}$ as a function of the effective system size $|\Omega_\epsilon(t)|$, with $\epsilon=0.001$.
3) and 4) Probability density $P_\theta(m)$ for $\theta=\tfrac{\pi}{2}$ and $\theta=\tfrac{\pi}{4}$ respectively.
}
\label{fig:example1}
\end{figure}

\begin{figure}[!t]
    \centering
    \includegraphics[width=.9\linewidth]{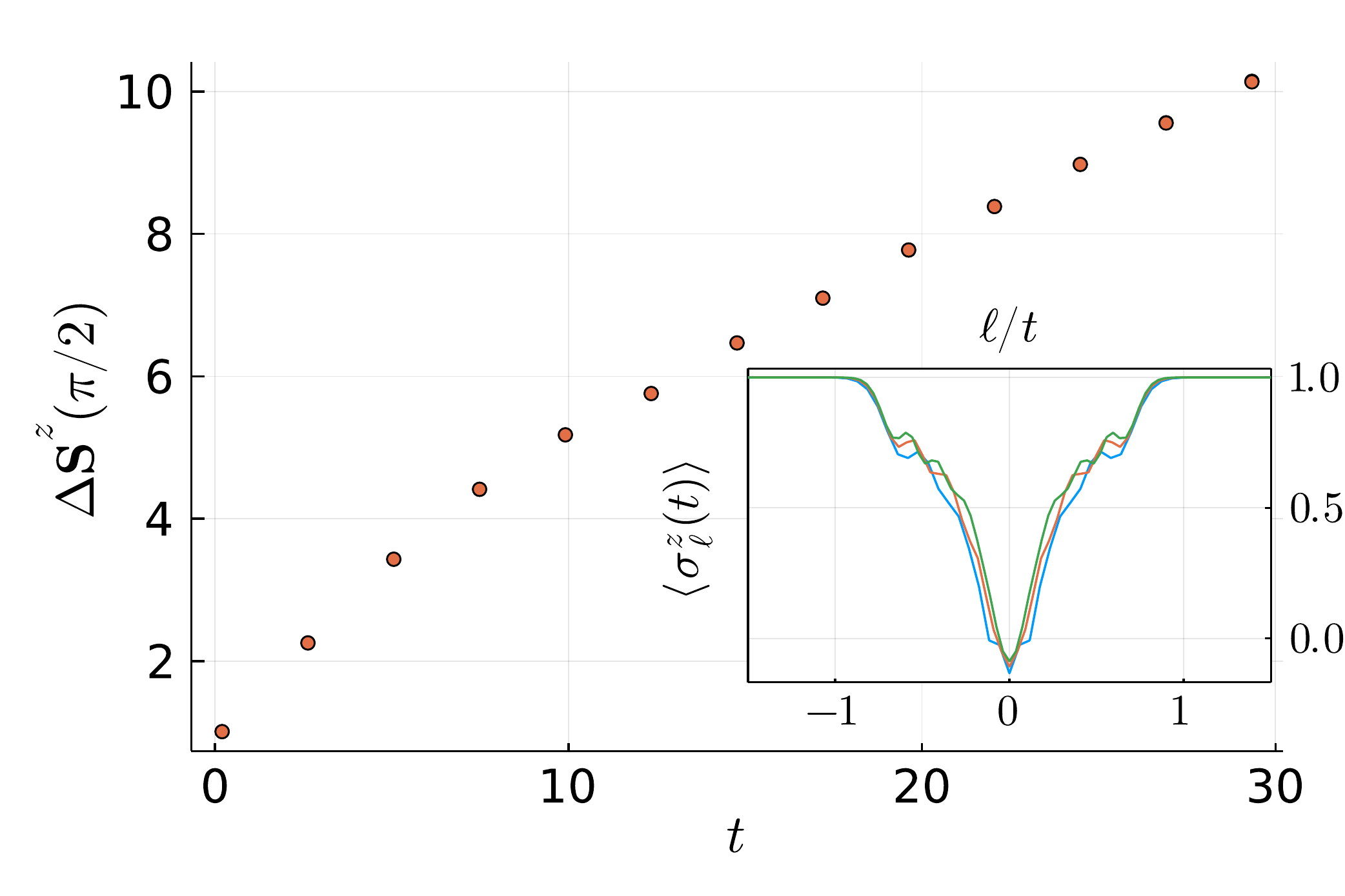}
    \includegraphics[width=.9\linewidth]{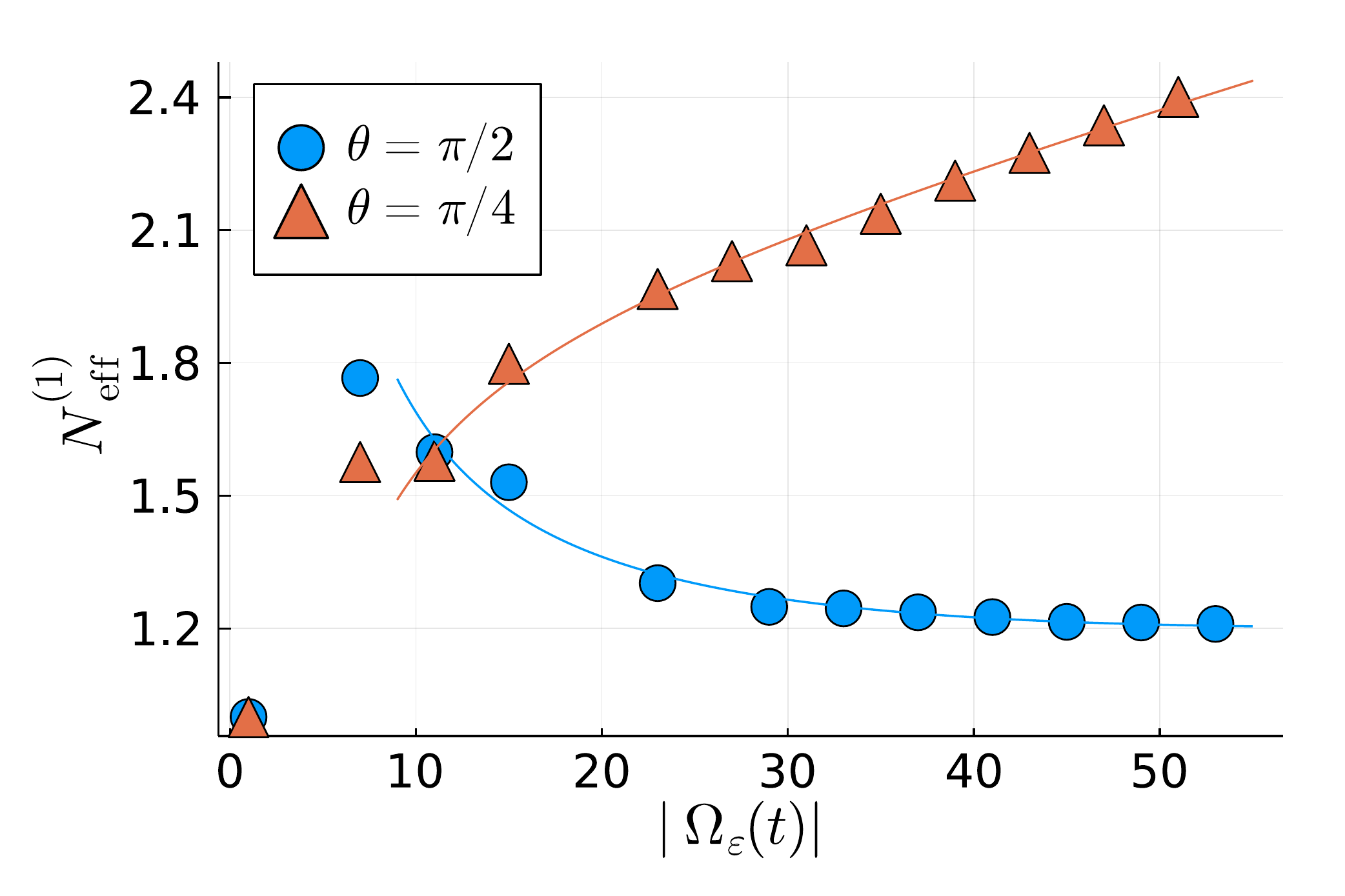}
    \caption{The same as in the top plots of Fig.~\ref{fig:example1} for Hamiltonian $\bs H_1$ with $J=1, \gamma=0.5, w=0.7, \Delta=0,D^z =0.6,h^z=0$. 
    }
    \label{fig:quantum_fisher_general}
\end{figure}

We measure macroscopic entanglement with the so called \textit{quantumness}, which characterises the asymptotic behaviour of the maximum quantum Fisher information among all extensive observables (see \textit{Materials and Methods}). 
In fact, we only compute a lower bound of it, $\mathcal{N}_{\mathrm{eff}}^{(1)}$, resulting from reducing the space of observables.  
An example is reported in Fig.~\ref{fig:example1}, where we plot  $\mathcal{N}_{\mathrm{eff}}^{(1)}$ as a function of the effective system size $|\Omega_\epsilon(t)|$. To study the asymptotic behavior of $\mathcal{N}_{\mathrm{eff}}^{(1)}$ we fit the data with a curve parametrised as $\beta_0+\beta_1 |\Omega_\epsilon(t)|+\beta_2 \frac{1}{|\Omega_\epsilon(t)|}$, where the term $\frac{1}{|\Omega_\epsilon(t)|}$ stands for a potential sub-leading term. The estimated leading order is consistent with a constant ($\beta_1\approx0$) in the case $\theta=\pi/2$ and with a linear growth ($\beta_1\neq 0$) in the generic case.
The figure also shows how the difference of magnetization between flipping or not the spin is proportional to $|\Omega_\epsilon(t)|$, meaning that the two states $\ket{\Uparrow}$ and $\ket{\Psi_{\pi/2}(t)}$ are macroscopically different and therefore the state $\ket{\Psi_\theta(t)}$ is macroscopically entangled for generic $\theta$.
Fig.~\ref{fig:example1} also reports the probability density $P_\theta(m)$ to get $m$ from a measurement of $\bs S^z$ given that the system is in the state $\ket{\Psi_\theta(t)}$; the case $\theta=\frac{\pi}{2}$ shows standard fluctuations, while the more general case $\theta=\frac{\pi}{4}$ shows the bimodal distribution that is characteristic of a cat state. 
We have shifted $m$ by the expectation value of $\bs S^z(t)$ so as to exhibit plots that are independent of the system's size (incidentally, this also makes it equivalent to consider either the full chain or $\Omega_\epsilon(t)$).
The probability distribution in the GHZ state $(\ket{\Uparrow}+\ket{\Downarrow})/\sqrt{2}$ would exhibit two Kronecker deltas at the maximum and minimum magnetisations; here one delta is replaced by a Gaussian centred at a different magnetization, but it is still well separated from the other peak. 
We have checked that, as long as the initial state is in the bulk of the spectrum, the qualitative behaviour remains the same even moving away from integrability. An example is reported in Fig.~\ref{fig:quantum_fisher_general}.
Remarkably, even in this non-integrable case the profiles of local magnetization (the insets in the plots of the total magnetization) computed at different times collapse to the same curve in the ray coordinate $\ell/t$, which manifests the ballistic change in the total magnetization.
Note also that from the profile of the local magnetization we get a visual definition of $\Omega_\epsilon(t)$: it essentially coincides with the region with $\braket{\sigma^z_\ell(t)}\not\approx 1$.
We mention that the tidy structure of the transistor Hamiltonian $\bs H_1$ allows for an alternative interpretation of the protocol. 
The state before the measurement has all transistor's \textit{switches} open, and time evolution is blocked. 
The measurement has the effect of turning a switch in a quantum superposition of open and closed (the switch is exactly closed only for $\theta=\frac{\pi}{2}$).  
Closing a switch enables Heisenberg and Dzyaloshinskii-Moriya interactions between the neighboring spins of the opposite leg of the chain, which, in turn, partially close other switches, and so on and so forth, 
leading to a complete reconfiguration of the state that affects a region whose size grows linearly in time.

\subsection{Generic model}
\begin{figure}[!t]
    \centering
    \includegraphics[width=0.9\linewidth]{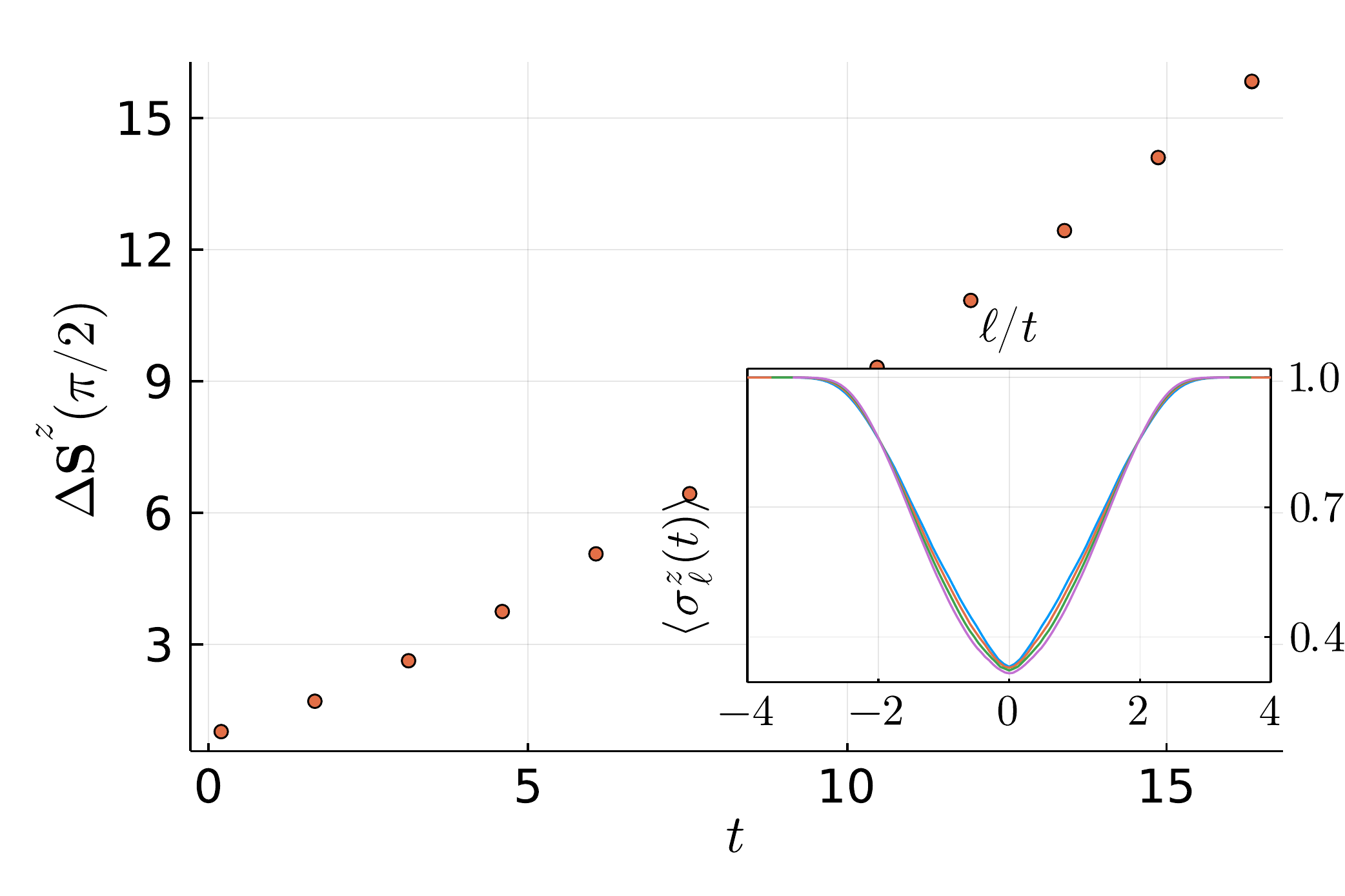}
    \includegraphics[width=0.9\linewidth]{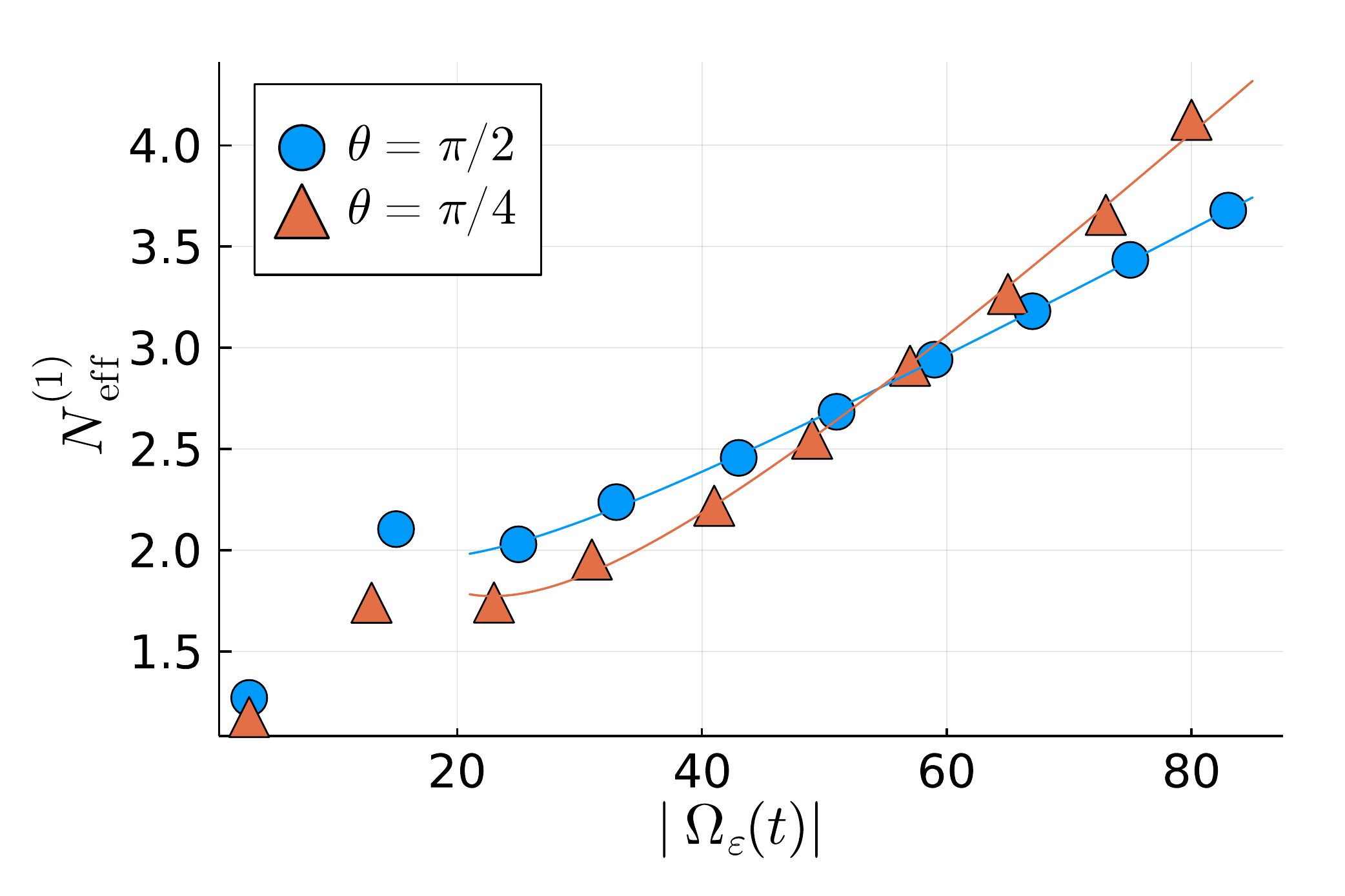}
    \caption{The same as in the top plots of Fig.~\ref{fig:example1} for Hamiltonian $\bs H_2$ with $J_1=J_2=1$, $J_{r>2}=0$, $\gamma^z_1=\gamma^z_2 = -0.6$, $D_{r>1}^y=h^z=\gamma^z_{r>2}=0$, $D_1^y = -0.9$, $\gamma^x_r=\gamma^y_r=D^x_r=D^z_r=0,\forall r$. 
    The widest profile in the inset corresponds to the largest time.
    } 
    \label{fig:fcs_longrange}
\end{figure}

We have considered several models without a hidden $U(1)$ symmetry in which $\ket{\Uparrow}$ is still a quantum scar.
We see quite generally that, also in those cases, the spin flip has everlasting effects arbitrarily far from the origin. 
In Fig.~\ref{fig:fcs_longrange} we report an example using the Hamiltonian $\bs H_2$. The plot of the magnetization shows that the states $\ket{\Uparrow}$ and $\ket{\Psi_{\pi/2}(t)}$ are macroscopically different, indeed the difference of the magnetization in the two states grows linearly with the effective system size. Again, we conclude that our protocol leads to the formation of a macroscopically entangled state. Concerning the  quantumness, we have fitted the data for the lower bound $\mathcal{N}_{\mathrm{eff}}^{(1)}$ with the same Ansatz as before $\beta_0+\beta_1 |\Omega_\epsilon(t)|+\beta_2 \frac{1}{|\Omega_\epsilon(t)|}$; the analysis points to an asymptotic linear growth for any $\theta\neq 0$, including this time also $\theta=\frac{\pi}{2}$. 
Therefore, $\ket{\Psi_{\theta}(t)}$ still exhibits macroscopic entanglement for generic $\theta$, but, in contrast to the quantum transistor chain, also $\ket{\Psi_{\pi/2}(t)}$ is macroscopically entangled, undermining in turn the formation of a cat state.
%
\subsection{Imperfections in the preparation of the state}
We investigate the effect of perturbing the state by time evolution under a homogeneous Hamiltonian $\bs H_0$ for a fixed time $t_0$, i.e., $\ket{\Psi_0(0^-)}= e^{-i\bs H_{0} t_0}\ket{\Uparrow}$.
We then study the time evolution of $\ket{\Psi_{\theta}(0^+)}=e^{i\theta \bs \sigma_0^y}\ket{\Psi_0(0^-)}$, which is the analog of Eq.~[\ref{eq:Psi0}].
We warn the reader of a complication:  the initial evolution affects the whole chain, so, strictly speaking, there is no approximately pure subsystem.
If we insist on considering the subsystem associated with the spreading of the local perturbation (corresponding to the sharp change of behaviour in the profile of the magnetization), we should analyse the quantum Fisher information of a mixed state, which is a harder quantity to compute.
In this preliminary study of the stability of our protocol, we limit ourselves to check the qualitative behaviour of the probability distribution of the magnetization. 

We distinguish two classes of perturbations: in the first class, the perturbation is spin-flip symmetric (thus the hidden $U(1)$ symmetry is preserved), e.g., $\bs H_0$ could be the integrable Ising Hamiltonian 
\begin{equation}
\bs H_{0,1}=-  \frac{1}{4}\sum_\ell \left( \bs\sigma_\ell^x\bs\sigma_{\ell+1}^x+h_0^z\bs\sigma_\ell^z\right)\, ;
\end{equation}
in the second class, the pre-measuerement Hamiltonian breaks the spin-flip invariance behind the hidden $U(1)$ symmetry, e.g.
\begin{equation}
\bs H_{0,2}=-\frac{1}{4}\sum_\ell\left(  \bs\sigma_\ell^x\bs\sigma_{\ell+1}^x+h_0^z\bs\sigma_\ell^z+h_0^x\bs\sigma_\ell^x\right)\,,
\end{equation}
which is a non-integrable version of the Ising model.
Note that, in both cases, the larger $h_0^z$, the smaller the expected effect, therefore $h_0^z$ can be used as a control parameter.

The first class of perturbations is not expected to destabilize the growth of a cat state: On the one hand,  Ref.~\cite{Fagotti2022Global} showed that the perturbation has everlasting macroscopic effects even after so-called ``global quenches'' from a spin-flip invariant initial state. On the other hand, Ref.~\cite{Maric2022Universality} indirectly confirms clustering in $\ket{\Psi_{\pi/2}(t)}$ at long times with noninteracting transistor Hamiltonians, indeed the mutual information approaches zero at large distances.  
Our numerical analysis is consistent with these expectations, indeed we still see the formation of a bimodal probability distribution (with two well separated peaks). 

\begin{figure}[!t]
    \centering
    \includegraphics[width=.9\linewidth]{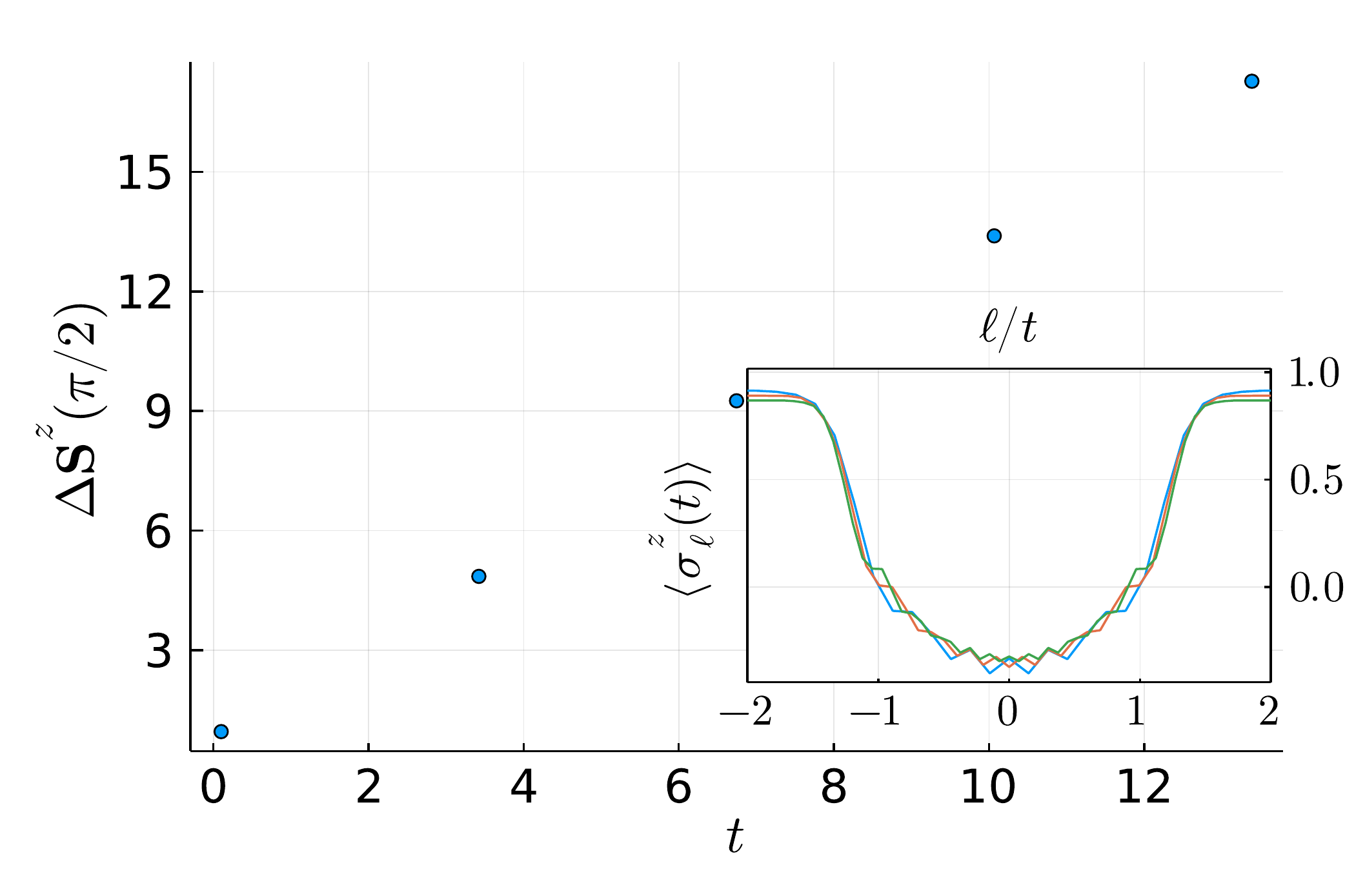}
    \includegraphics[width=.9\linewidth]{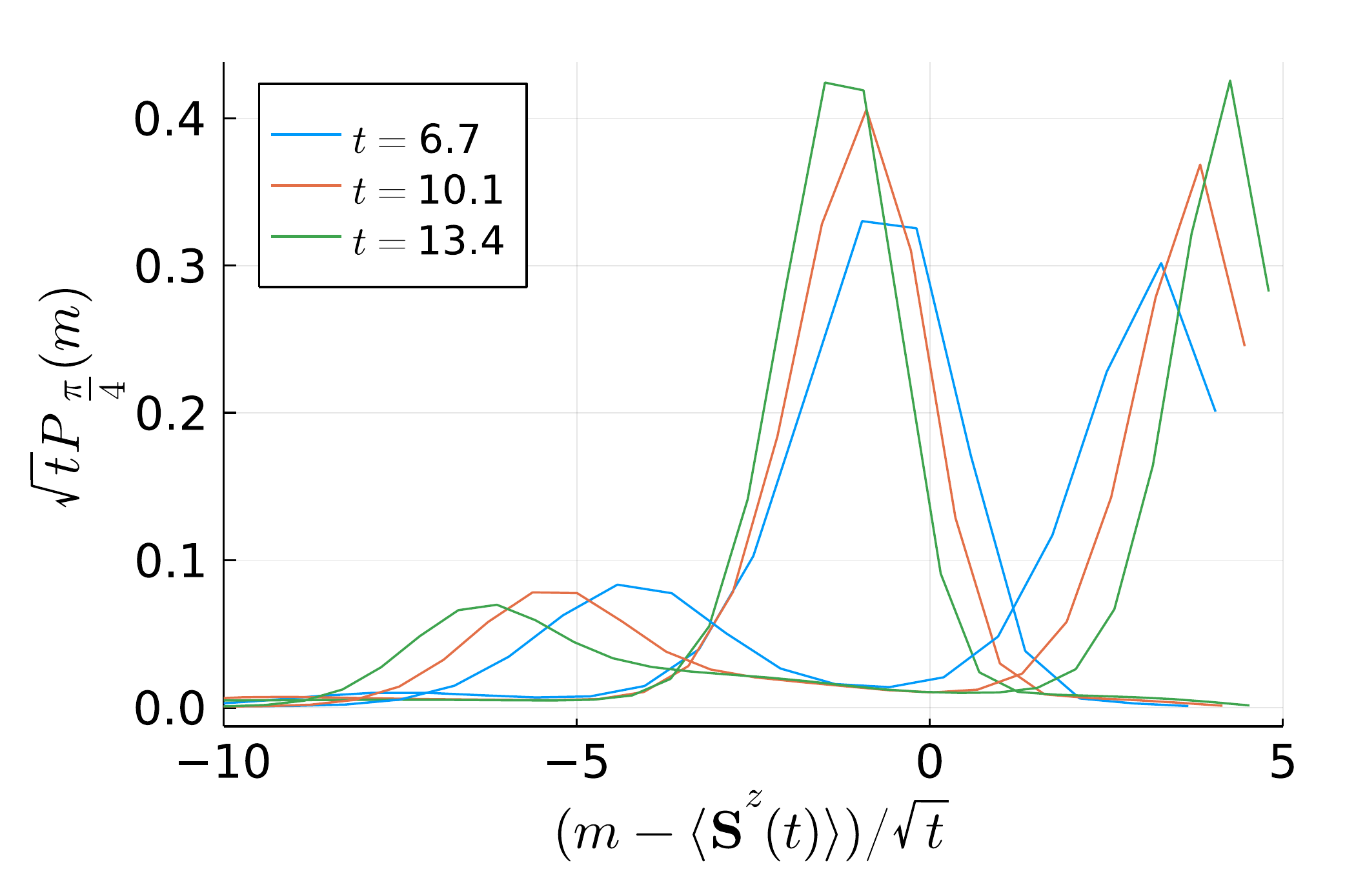}
    \caption{Time evolution under $\bs H_1$ after perturbing the state $\ket{\Uparrow}$ with  $\bs H_{0,2}$ for a time $t_0=0.4$. 
    The parameters are set to $J=2.8$, $\gamma=1$, $w=\Delta=D^z=h^z=0$, for $\bs H_1$,  and $h^z_0=0.7$ and $h^x_0=0.5$ for $\bs H_{0,2}$.
    Top: the difference in magnetization with or without the spin flip grows linearly in time, at least for the times reached by our simulation.
    Bottom: more than two peaks appear and they are not well separated. 
    }
    \label{fig:dirty_fcs2}
\end{figure}

On the contrary, for the second class of perturbations the probability density does not present two well separated peaks, as shown e.g. in the bottom row of Fig.~\ref{fig:dirty_fcs2}. 
This rules out the formation of a cat state, but  the basic features pointing to macroscopic entanglement (such as the linear growth in time of the difference of magnetization between flipping or not the spin) are still present for times longer than our maximum simulation times.

\section{Discussion}
\subsection{Local perturbations with macroscopic effects}
In ordinary 1D systems a spin flip does not change the macroscopic properties of a state whatever long the time is. The naive expectation is that its effect spreads out across the chain as $t^\alpha$ but fades away as $t^{-\alpha}$, with $\alpha=1$ in integrable models and $\alpha=\tfrac{1}{2}$ in most of the generic systems. 
Recently, three situations have been pointed out in which this picture can break down:
\begin{itemize}
\item ground states in symmetry-broken phases~\cite{Zauner2015,Eisler2020Front}. 
\item quantum jammed states in systems with Hilbert space fragmentation~\cite{Bidzhiev2022Macroscopic}.
\item systems with semilocal conservation laws~\cite{Fagotti2022Global}. 
\end{itemize}
To these three known cases, we have added models with a fully separable quantum scar. 
This novel scenario has two advantages: first, a product state can be easily prepared experimentally, and, second, its stationarity
gives the freedom to stop the (virtual) experiment at any time without compromising the outcome. 

Remarkably, the norm is that the perturbation generates macroscopic entanglement, indeed, the possible but improbable operators breaking that rule can be cured by the addition of a local operator that does not change the excited state --- e.g., $\epsilon \bs\sigma_\ell^z$ for our initial state $\ket{\Uparrow}$.

In the special case in which there is a hidden $U(1)$ symmetry our analysis of the fluctuations reveals that the entangled part of the system is in a Schr\"odinger's cat state. 

\subsection{On the spreading}
The existence of the semilocal charge $\sum_\ell\bs\Pi^z(\ell)$ ensures that the effect of the spin flip does not fade away at long time~\cite{Fagotti2022Global,Fagotti2022Nonequilibrium}, but it is not sufficient to exclude the effect to remain confined around the position of the measurement. This is indeed what happens when $h$ is too large, in which case the Hamiltonian exhibits localised excitations (see Appendix~\ref{app:localised_excit}). In the generic case, however, the perturbation spreads out. Using the underlying integrable structure, Ref.~\cite{Fagotti2022Global} showed that the spreading is ballistic for $\gamma=1$, $w=\Delta=D^z=0$, and $|h^z|<\frac{1}{4}$. And for slightly different initial product states,  ballistic behaviour was observed~\cite{Bidzhiev2022Macroscopic,Zadnik2021Measurement}  also for $\gamma=w=\Delta=D^z=h^z=0$, for which the model is integrable as well. 
A priori one would not expect ballistic spreading in generic models, but rather diffusive. In Ref.~\cite{Bidzhiev2022Macroscopic}, in particular, a preliminary analysis of the effect of some integrability-breaking perturbations pointed to a non-ballistic spreading. In contrast, we find numerical evidence that ballistic behaviour extends to a wide range of parameters, at least, within the time window investigated. In order to rule integrability out, we have checked the statistics of the energy levels, and, for  generic parameters, our data are in excellent agreement with a Wigner-Dyson distribution (see the Appendix~\ref{app:energy_statistics}), supporting, in turn, the non-integrability of the generic models investigated. 

\section{Materials and Methods}

\subsection{Simulations of the dynamics}
Numerical simulations are performed with Julia ITensor library \cite{itensor}.
We use a time-evolving block decimation (TEBD) algorithm and, in all data reported, time evolution is discretized in time steps $\delta t = 0.01$ with 2nd order Trotter-Suzuki gates \cite{Hatano2005}. 

\subsection{Full counting statistics}
The full probability distribution of a given observable encodes the entire information about its moments.
Let $L$ be the chain's length.
We define $P_\theta(m)$ as the probability to get $m$ from a measurement of $\bs S^z$ given that the system is in the state $\ket{\Psi_\theta(t)}$.
Note that $m\in\{-\frac{L}{2}, -\frac{L}{2}+1,...,\frac{L}{2}\}$.
To set a convention, we assume $L$ to be divisible by $4$, so that the maximum possible value $m$ of the magnetization is an integer even number.
The generating function of the moments of the probability distribution is defined as
\begin{equation}
G_\theta(k) = \braket{\Psi_\theta(t)| e^{i \frac{2\pi k}{L+1} \bs S^z}|\Psi_\theta(t)}.
\end{equation}
The probability $P_\theta(m)$ is then the Fourier transform of the generating function:
\begin{equation}
P_\theta(m) 
= \frac{1}{L+1} \sum_{k=-L/2}^{L/2} e^{-i \frac{2\pi k}{L+1} m} G_\theta(k).
\end{equation}
We compute the generating function numerically using Julia ITensor library \cite{itensor}.

We point out that, if the model is invariant under spin flip $\bs\sigma^{x,y}\rightarrow -\bs \sigma^{x,y}$, the state $\ket{\Psi_{\pi/2}(t)}$ belongs to the sector in which the parity operator $\bs\Pi^z=\prod_{\ell}\bs\sigma^z_\ell$ has eigenvalue $-1$, which means $P_{\pi/2}(m)=0$ for any even $m$.
This implies that, in the generic case of $\ket{\Psi_\theta(t)}$ with $\theta\neq\frac{\pi}{2}$, $P_\theta(m)=0$ for any even $m$ except for $P_\theta(L/2)=\cos^2\theta$. 
This is why, in the plots of the probability density for models with spin-flip invariance, we report only odd values of $m$ and $m=L/2$.

\subsection{Quantumness and Quantum Fisher information}
Ref.~\cite{smerzi2009} proposed the quantum Fisher information as a measure of the ``macroscopicity'' of quantum effects in lattice systems.
Its computation is however rather complicated when addressed in full generality, both analytically and numerically. 
For the sake of simplicity, we follow the approach of Refs~\cite{Frowis2012Measures,Hyllus2012Fisher,Toth2012} and restrict ourselves to examining a sufficient condition. Specifically, we investigate the quantum Fisher information $\mathcal F(\bs O)$ of extensive operators whose densities have support on a single site, i.e., $\bs O[\{\vec n \}]=\sum_{j\in\Omega_\epsilon(t)}\vec n_j\cdot\vec{\bs\sigma}_j$, where $\vec{\bs\sigma}_j\equiv \{\bs\sigma^x_j,\bs\sigma^y_j,\bs\sigma^z_j\}$ and the coefficients are normalised as $|\vec n_j|^2=1$.
In pure states, which is the case we consider, $\mathcal F(\bs O)$ equals four times the variance of $\bs O$. 
Using the notation of \cite{Frowis2018}, we introduce the quantumness $N_{\mathrm{eff}}$ of a state as its maximal quantum Fisher information with respect to all extensive observables:
\begin{equation}
    N_{\mathrm{eff}}=\frac{1}{4|\Omega_\epsilon(t)|}\max_{\bs O}\mathcal{F}(\bs O).
\end{equation}
We denote by $N_{\mathrm{eff}}^{(1)}$ the maximisation restricted to the observables with single-site density as introduced above.
$N_{\mathrm{eff}}$ was interpreted as an effective size of the macroscopic quantum state, thus $N_{\mathrm{eff}}^{(1)}$ is a lower bound for the effective size. 
We introduce the covariance matrix 
\begin{multline}
[\mathrm K(t)]_{n,\alpha;m,\beta}=\frac{1}{2}\braket{\Psi_{\pi/2}(t)|\{\bs\sigma_n^\alpha, \bs\sigma_m^\beta\}|\Psi_{\pi/2}(t)}
+\\-\braket{\Psi_{\pi/2}(t)|\bs\sigma_n^\alpha|\Psi_{\pi/2}(t)}\braket{\Psi_{\pi/2}(t) |\bs\sigma_m^\beta|\Psi_{\pi/2}(t)}\,,
\end{multline} 
where $\{\bs\sigma_n^\alpha, \bs\sigma_m^\beta\}=\bs\sigma_n^\alpha \bs\sigma_m^\beta+\bs\sigma_m^\beta\bs\sigma_n^\alpha$.
In pure states, $\mathcal N_{\mathrm{eff}}^{(1)}$ satisfies
\begin{equation}\label{eq:L1}
\mathcal N_{\mathrm{eff}}^{(1)}=\frac{\mathrm{tr}[\mathrm D(t)]}{|\Omega_\epsilon(t)|}\, ,\quad\text{with}\quad\mathrm K(t)\vec v(t)=[\mathrm D(t)\otimes \mathrm I_3] \vec v(t)
\, ,
\end{equation}
where $n,m\in\Omega_{\epsilon}(t)$ (the vector space is $3|\Omega_{\epsilon}(t)|$-dimensional) and $\mathrm D(t)$ is required to be diagonal ($[\vec v]_n$ are normalised). 
We solve Eq.~[\ref{eq:L1}] using a Lanczos algorithm with three basic iterative steps:
\begin{enumerate}
\item $\vec w^{(n)}=\mathrm{K}\vec v^{(n)}$
\item $[\mathrm D^{(n)}]_j=\parallel\vec w_j^{(n)}\parallel^{-1}$
\item $\vec v^{(n+1)}=(\mathrm D^{(n)}\otimes \mathrm I_3)\vec w^{(n)}$
\end{enumerate}
In most of the cases this procedure worked well without particular stabilizers.  

\section{Conclusion}
We have shown that, quite generally, macroscopic entanglement emerges from a localised inhomogeneity in a state that would be otherwise stationary. This can not happen in an ordinary excited state that maximises the entropy under the constraints of the local conservation laws. In generic systems, indeed, we observe the phenomenon in quantum scars with anomalously low bipartite entanglement. We also argue that, in the specific case in which the excited state maximises instead the entropy under the constraints of also semi-local conservation laws, macroscopic entanglement takes the simple form of a cat state, providing in turn a novel way to generate cat states in a class of quantum many-body systems with local interactions. 

This work leaves several open questions. First, a rigorous proof of the generation of cat states is missing. Second, the observation of ballistic behaviour in generic systems could be questioned as a finite-time effect, so additional investigations are imperative. Third, we have only provided a preliminary and incomplete check of the stability of our protocol, but this is clearly an important  issue and does require additional investigations. 

\begin{acknowledgments}
S.B. thanks Tommaso Roscilde and Augusto Smerzi for useful discussions. This work was supported by the European Research Council under the Starting Grant No. 805252 LoCoMacro.
\end{acknowledgments}

\bibliographystyle{unsrtnat}
\bibliography{biblio.bib}

\onecolumn\newpage
\appendix

\section{Energy level statistics}
\label{app:energy_statistics}

As a test of integrability or otherwise, we investigate the energy level statistics of the Hamiltonian of the spin-$\frac{1}{2}$ nearest-neighbor transistor chain
\begin{equation}
\bs H=\sum_\ell \frac{\bs 1-\bs \sigma_\ell^z}{8}[J\vec{\bs \sigma}_{\ell-1}\overset{S}{\cdot} \vec{\bs \sigma}_{\ell+1}+\vec D\cdot (\vec{\bs \sigma}_{\ell-1}\times \vec{\bs \sigma}_{\ell+1})]-\frac{\vec h}{2}\cdot\vec{\bs\sigma}_\ell,
\end{equation}
where $\vec a\overset{S}{\cdot}\vec b=\vec a\cdot( S\vec b)$ and
\begin{equation}
S=\begin{bmatrix}
\frac{1+\gamma}{2}&w&0\\
w&\frac{1-\gamma}{2}&0\\
0&0&\Delta
\end{bmatrix}\quad \vec D=\begin{bmatrix}
0\\
0\\
D^z
\end{bmatrix}\quad \vec h=\begin{bmatrix}
0\\
0\\
h^z
\end{bmatrix}\, .
\end{equation}
We enforce periodic boundary conditions and restrict ourselves to the sector characterised by zero momentum, zero semilocal charge $\tilde{\bs S}^z$ (the largest eigenspace), and  $\prod_{j}\bs\sigma_j^z=\prod_{j}\bs\sigma_{2j}^z=1$. 

\begin{figure}[h]
    \centering
    \includegraphics[width=.6\textwidth]{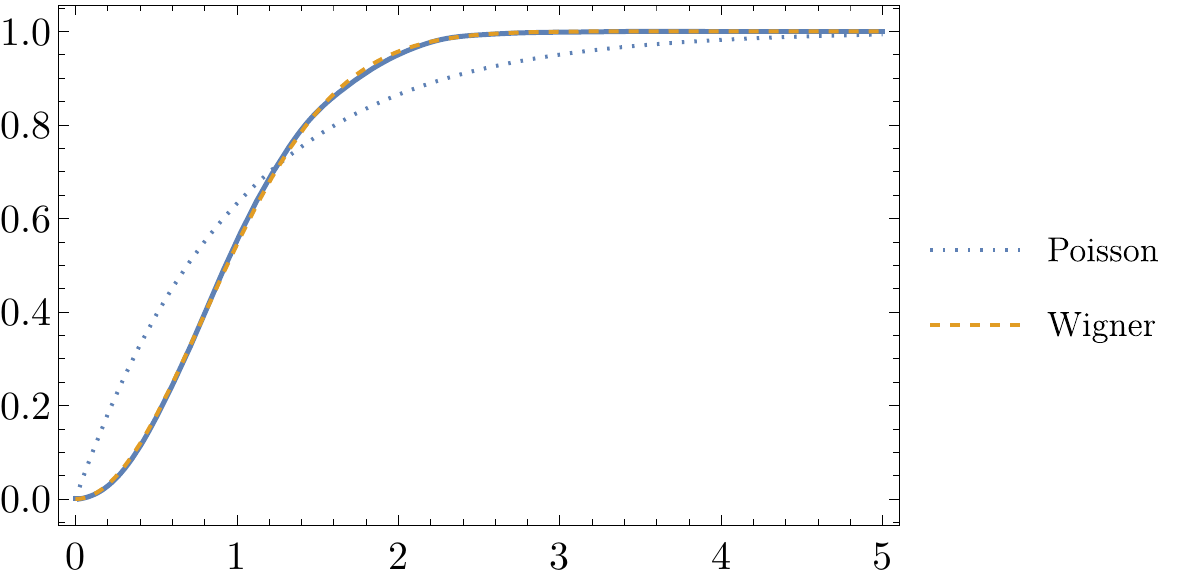}
    \caption{
    Cumulative distribution of the unfolded energy level spacing in a sector of $\bs H$ with $J=1, \gamma=0.5, w=0.7, \Delta=0,D^z =0.6,h^z=0$ in a chain with $20$ spins. The data are shown as a solid curve; the dotted and dashed curves are the (normalised) Poisson and Wigner-Dyson predictions expected in  integrable and generic models, respectively.
    }
    \label{fig:SI}
\end{figure}

Such a sector is equivalent to the sector with zero momentum, zero magnetization, and $\prod_{j}\bs\tau_j^x=\prod_{j}\bs\tau_j^z=1$,  of the Hamiltonian
\begin{multline}
\bs H_\tau=\sum_\ell J\frac{(1+\gamma)\bs I+(1-\gamma)\bs\tau_{\ell-1}^z\bs\tau_{\ell+2}^z}{4}\frac{\bs\tau_{\ell}^x\bs\tau_{\ell+1}^x+\bs\tau_{\ell}^y\bs\tau_{\ell+1}^y}{4}
+\\+\frac{(D^z+J w)\bs\tau_{\ell+2}^z+(D^z-J w)\bs\tau_{\ell-1}^z}{2}\frac{\bs\tau_\ell^x\bs\tau_{\ell+1}^y-\bs\tau_\ell^y\bs\tau_{\ell+1}^x}{4}-\frac{h^z}{2}\bs\tau_{\ell}^z\bs\tau_{\ell+1}^z
\end{multline}
with periodic boundary conditions. Indeed, $\bs H_\tau$ is related to $\bs H$ by a Kramers-Wannier duality mapping which transforms  semilocal operators into odd local ones~\cite{Fagotti2022Global}
\begin{equation}
\label{eq:finite_duality}
\bs\sigma_j^x=\bs\Pi_{\tau,-}^x(j) \qquad\quad
\bs\sigma_j^y=\begin{cases}
\bs \tau_1^x\bs \tau_2^z&j=1\\
\bs\Pi^x_{\tau,-}(j-1)\bs\tau_j^y\bs\tau_{j+1}^z&1<j<L\\
-\bs\tau_L^z&j=L
\end{cases}\qquad\quad
\bs\sigma_j^z=\begin{cases}
\bs\tau_j^z\bs\tau_{j+1}^z&1\leq j<L\\
\bs\Pi^x_\tau\bs\tau_1^z\bs\tau_L^z&j=L\,,
\end{cases}
\end{equation}
where
\begin{align}
\bs\Pi^x_{\tau,-}(j)=-\bs\tau_1^y\prod_{\ell=2}^{j}\bs\tau_\ell^x,\qquad
\bs\Pi^x_{\tau}=\prod_{\ell=1}^{L}\bs\tau_\ell^x\, .
\end{align}
Fig.~\ref{fig:SI} shows quite clearly that the Hamiltonian of Fig.~3 in the main text is not integrable. Indeed, the cumulative distribution of the energy level spacing is in excellent agreement with the Wigner-Dyson distribution.

\section{Localised excitations}\label{app:localised_excit}
\begin{figure}[h]
    \centering
\includegraphics[width=.5\textwidth]{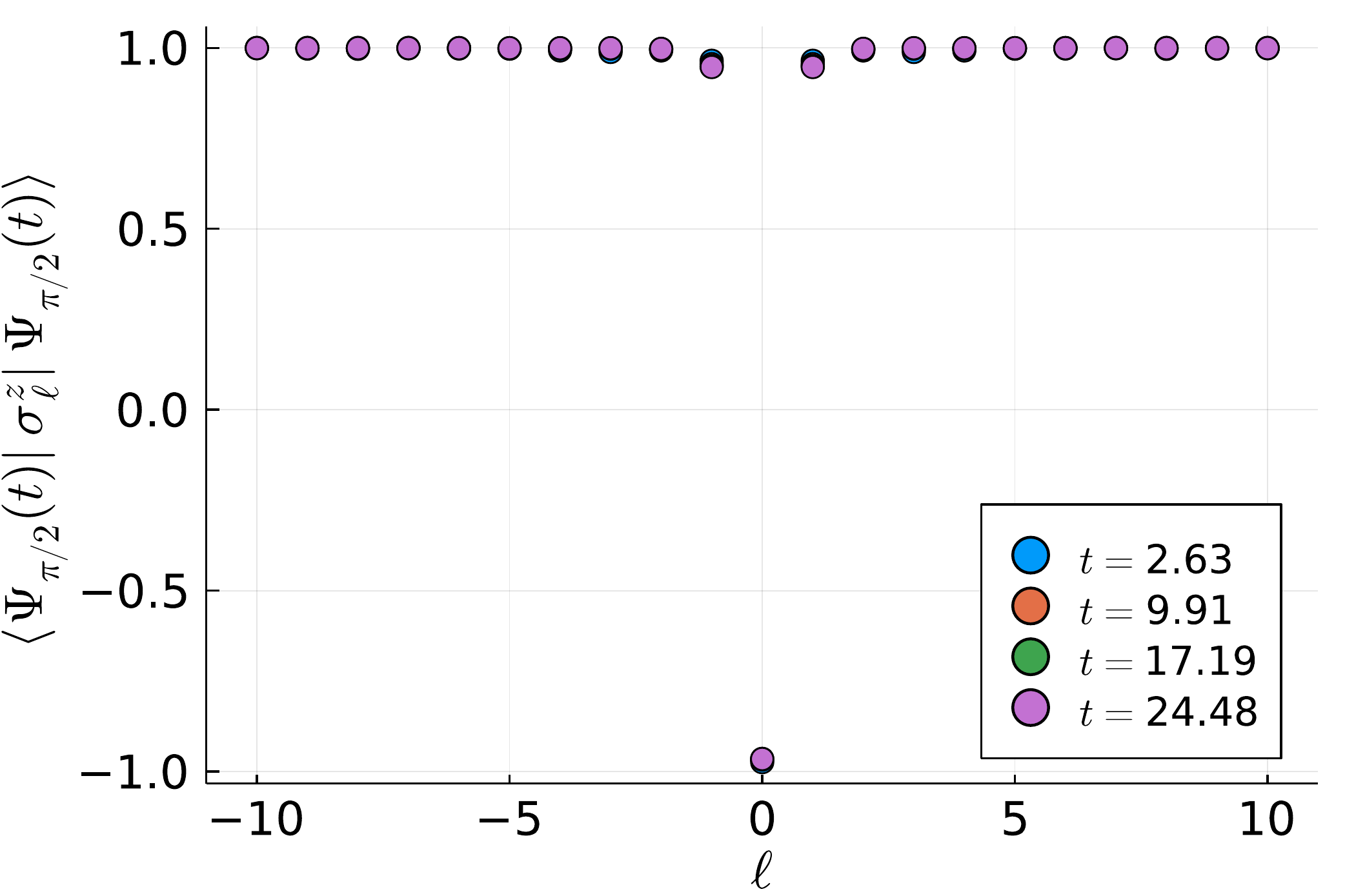}
    \caption{The magnetization does not change for any accessible time. The fact that the magnetization is maximal except for a few sites around the origin tells us that most of the system is in the $\ket{\Uparrow}$ state.}
    \label{fig:spike}
\end{figure}

The main exception to the development of macroscopic entanglement after a local projective measurement is when the pre-measurement state is the ground state in a disordered phase.  
We consider, for example, the dual XXZ Hamiltonian $H=\sum_\ell \bs\sigma^x_{\ell-1}\frac{1-\bs\sigma^z_\ell}{8}\bs\sigma^x_{\ell+1}-\frac{h^z}{2}\bs\sigma^z_\ell$. 
If the intensity of the magnetic field $|h^z|$ is larger than $\frac{1}{2}$, the pre-measurement state becomes the ground state, and
a spin flip excites only local modes that do not spread with time~---~see Fig.~\ref{fig:spike}.
%
\section{Absence of macroscopic effects in models with a $U(1)$ symmetry}\label{app:U(1)}
In this section we prove that a state $\ket{\Psi}$  obtained by time evolution under a $U(1)$-symmetric Hamiltonian after a local perturbation to $\ket{\Uparrow}$
is macroscopically equivalent to $\ket{\Uparrow}$; that is to say, the quantity $\braket{\Psi|\bs O|\Psi}-\braket{\Uparrow|\bs O|\Uparrow}$ is subextensive for any translational-invariant operator $\bs O = \sum_\ell \bs O_{A_\ell}$, with $\bs O_{A_\ell}$ local operators.
We also prove that the variance of any of such operators $\bs O$ with respect to the state $\ket{\Psi}$ grows at most linearly with system  $|\Omega_\epsilon(t)|$, ruling out the possibility to obtain a macroscopically entangled state from time-evolution of a locally-perturbed $\ket{\Uparrow}$.

Let us denote by $\ket{\Psi_s}$ a state obtained by flipping $s$ spins in $\ket{\Uparrow}$, or linear combination of them.
We start by observing that, if $\bs S^z$ is conserved, the state time-evolves as 
$$\ket{\Psi_{s}(t)}=
\sum_{n_1<n_2<...<n_s} w_{\{n\}}(t) \bs\sigma_{n_1}^-\bs\sigma_{n_2}^-...\bs\sigma_{n_s}^-\ket{\Uparrow} \,,
$$
for some coefficients such that $\sum_{n_1<n_2<...<n_s} |w_{\{n\}}|^2=1$.
Let $\bs O_{A_\ell}$ be an operator with support in $A$, where $A_\ell$ is a set of $|A|$ adjacent sites starting with the $\ell$-th spin: $A=\{\ell,...,\ell+|A|-1\}$. We have
{\small
\begin{multline}
\braket{\Psi_{s}(t)|\bs O_{A_\ell}|\Psi_{s}(t)}-
\braket{\Uparrow|\bs O_{A_\ell}|\Uparrow}
\\=\sum_{i=1}^{s}
\sum_{\substack{n_1<...<n_s\\n_k\in A_\ell \Leftrightarrow k=i}}
\biggl(\sum_{\substack{n_i'\\n_k'\in A_\ell,\forall k\in\{i\}}}
w_{\{n\}}^*(t)
w_{\{n\},n_i\rightarrow n'_i}(t)
\braket{\Uparrow|\bs\sigma^+_{n'_i}\bs O_{A_\ell}\bs\sigma^-_{n_{i}}|\Uparrow}
- 
|w_{\{n\}}(t)|^2 
\braket{\Uparrow|\bs O_{A_\ell}|\Uparrow}
\biggr)  
\\+
\sum_{i=1}^{s-1}
\sum_{\substack{n_1<...<n_s\\n_k\in A_\ell\Leftrightarrow k\in\{i,i+1\}}}
\biggl(\sum_{\substack{n_i'<n'_{i+1}\\n'_k\in A_\ell,\forall k\in\{i,i+1\}}}
w_{\{n\}}^*(t)
w_{\substack{\{n\},n_i\rightarrow n'_i\\n_{i+1}\rightarrow n'_{i+1}}}(t)
\braket{\Uparrow|\bs\sigma^+_{n'_i}\bs\sigma^+_{n'_{i+1}}\bs O_{A_\ell}\bs\sigma^-_{n_i}\bs\sigma^-_{n_{i+1}}|\Uparrow}
-
|w_{\{n\}}(t)|^2
\braket{\Uparrow|\bs O_{A_\ell}|\Uparrow}
\biggr)
\\+...+
\sum_{\substack{n_1<...<n_s\\ n_k\in A_\ell,\forall k\in\{1,...,s\}}}
\biggl(\sum_{\substack{n'_1<...<n'_s\\ n'_k\in A_\ell,\forall k\in\{1,...,s\}}}
w_{\{n\}}^*(t)
w_{\{n'\}}(t)
\braket{\Uparrow|\bs\sigma^+_{n_1'}...\bs\sigma^+_{n_s'}\bs O_{A_\ell} \bs\sigma^-_{n_1}...\bs\sigma^-_{n_s}|\Uparrow}
- |w_{\{n\}}(t)|^2
\braket{\Uparrow|\bs O_{A_\ell}|\Uparrow}
\biggr)\,.
\end{multline}}
An upper bound for the absolute value of the left hand side of the equation is obtained using $\braket{v|\bs O|v}\leq\parallel\bs O\parallel\braket{v|v}$, where $\parallel\bs O\parallel $ denotes the operator norm, indeed we have
\begin{multline}
|\braket{\Psi_{s}(t)|\bs O_{A_\ell}|\Psi_{s}(t)}-
\braket{\Uparrow|\bs O_{A_\ell}|\Uparrow}|
\leq\\\leq
2\parallel\bs O_{A_\ell}\parallel\left( \sum_{i=1}^{s}
\sum_{\substack{n_1<...<n_s\\n_k\in A_\ell \Leftrightarrow k=i}}
|w_{\{n\}}(t)|^2
+
\sum_{i=1}^{s-1}
\sum_{\substack{n_1<...<n_s\\n_k\in A_\ell\Leftrightarrow k\in\{i,i+1\}}}
|w_{\{n\}}(t)|^2
+...+
\sum_{\substack{n_1<...<n_s\\ n_k\in A_\ell,\forall k\in\{1,...,s\}}}
|w_{\{n\}}(t)|^2\right)
\leq\\\leq
2\parallel\bs O_{A_\ell}\parallel \sum_{i=1}^{s}
\sum_{\substack{n_1<...<n_s\\n_i\in A_\ell}}
|w_{\{n\}}(t)|^2\,,
\end{multline}
where the last step is not an equality because there are some coefficients $w$ that are counted more than once in the last line (e.g., if we take $s=2$ and $A_\ell=\{\ell,\ell+1\}$, there is originally just one term $|w_{\ell,\ell+1}|^2$, coming from the sum $\sum_{\substack{n_1<...<n_s\\ n_k\in A_\ell,\forall k\in\{1,...,s\}}}$, but the sum $ \sum_{i=1}^{s}
\sum_{\substack{n_1<...<n_s\\n_i\in A_\ell}}$ has two of such terms, that we get for $i=\ell$ and $i=\ell+1$).
Let us now consider the corresponding translational-invariant extensive operator $\bs O=\sum_{\ell\in \Omega_\epsilon(t)}\bs O_{A_\ell}$, where here $\Omega_\epsilon$ has a generalised definition with respect to the main text, since we want it to include all the light cones steaming from each spin flip. This gives
\begin{multline}
\label{eq:expectation value s vs UP}
|\braket{\Psi_{s}(t)|\bs O|\Psi_{s}(t)}-
\braket{\Uparrow|\bs O|\Uparrow}|
\leq
\sum_{\ell\in\Omega_\epsilon(t)} |\braket{\Psi_{s}(t)|\bs O_{A_\ell}|\Psi_{s}(t)}-
\braket{\Uparrow|\bs O_{A_\ell}|\Uparrow}|
\leq\\\leq
2\parallel\bs O_{A_1}\parallel
\sum_{\ell\in\Omega_\epsilon(t)}
\sum_{i=1}^{s}
\sum_{\substack{n_1<...<n_s\\n_i\in A_\ell}}
|w_{\{n\}}(t)|^2
\lesssim
2\parallel\bs O_{A_1}\parallel|A|
\sum_{i=1}^{s}
\sum_{n_1<...<n_s}
|w_{\{n\}}(t)|^2
=2s\parallel\bs O_{A_1}\parallel|A|
\, ,
\end{multline}
where we used translational invariance and  $\lesssim$ rather than  $\leq$ because $\Omega_\epsilon(t)$ is only approximately pure. 
This shows that $\ket{\Psi_s(t)}$ and $\ket{\Uparrow}$ are macroscopically indistinguishable for any time $t$.

We want now to extend the result to the square of the operator. We follow the same steps:
{\small\begin{multline}
\braket{\Psi_{s}(t)|\bs O_{A_\ell}\bs O_{A_{\ell'}}|\Psi_{s}(t)}-
\braket{\Uparrow|\bs O_{A_\ell}\bs O_{A_{\ell'}}|\Uparrow}
\\=\sum_{i=1}^{s}
\sum_{\substack{n_1<...<n_s\\n_k\in A_\ell\cup A_{\ell'} \Leftrightarrow k=i}}
\biggl(\sum_{\substack{n_i'\\n_k'\in A_\ell\cup A_{\ell'},\forall k\in\{i\}}}
w_{\{n\}}^*(t)
w_{\{n\},n_i\rightarrow n'_i}(t)
\braket{\Uparrow|\bs\sigma^+_{n'_i}\bs O_{A_\ell}\bs O_{A_{\ell'}}\bs\sigma^-_{n_{i}}|\Uparrow}
- 
|w_{\{n\}}(t)|^2 
\braket{\Uparrow|\bs O_{A_\ell}\bs O_{A_{\ell'}}|\Uparrow}
\biggr)  
\\+
\sum_{i<j=1}^{s}
\sum_{\substack{n_1<...<n_s\\n_k\in A_\ell\cup A_{\ell'}\Leftrightarrow k\in\{i,j\}}}
\biggl(\sum_{\substack{n_i'<n'_{j}\\n'_k\in A_\ell\cup A_{\ell'},\forall k\in\{i,j\}}}
w_{\{n\}}^*(t)
w_{\substack{\{n\},n_i\rightarrow n'_i\\n_{j}\rightarrow n'_{j}}}(t)
\braket{\Uparrow|\bs\sigma^+_{n'_i}\bs\sigma^+_{n'_{j}}\bs O_{A_\ell}\bs O_{A_{\ell'}}\bs\sigma^-_{n_i}\bs\sigma^-_{n_{j}}|\Uparrow}
+\\-
|w_{\{n\}}(t)|^2
\braket{\Uparrow|\bs O_{A_\ell}\bs O_{A_{\ell'}}|\Uparrow}
\biggr)
+\dots+
\sum_{\substack{n_1<...<n_s\\ n_k\in A_\ell\cup A_{\ell'},\forall k\in\{1,...,s\}}}
\biggl(\sum_{\substack{n'_1<...<n'_s\\ n'_k\in A_\ell\cup A_{\ell'},\forall k\in\{1,...,s\}}}
w_{\{n\}}^*(t)
w_{\{n'\}}(t)
\times\\\times
\braket{\Uparrow|\bs\sigma^+_{n_1'}...\bs\sigma^+_{n_s'}\bs O_{A_\ell}\bs O_{A_{\ell'}} \bs\sigma^-_{n_1}...\bs\sigma^-_{n_s}|\Uparrow}
- 
|w_{\{n\}}(t)|^2
\braket{\Uparrow|\bs O_{A_\ell}\bs O_{A_{\ell'}}|\Uparrow}
\biggr)\,,
\end{multline}}
from which
\begin{multline}
|\braket{\Psi_{s}(t)|\bs O_{A_\ell}\bs O_{A_{\ell'}}|\Psi_{s}(t)}-
\braket{\Uparrow|\bs O_{A_\ell}\bs O_{A_{\ell'}}|\Uparrow}|
\leq 2\parallel \bs O_{A_\ell}\parallel^2 \sum_{i=1}^{s}
\sum_{\substack{n_1<...<n_s\\n_k\in A_\ell\cup A_{\ell'} \Leftrightarrow k=i}}
|w_{\{n\}}(t)|^2 
+\\+
2\parallel \bs O_{A_\ell}\parallel^2\sum_{i<j=1}^{s}
\sum_{\substack{n_1<...<n_s\\n_k\in A_\ell\cup A_{\ell'}\Leftrightarrow k\in\{i,j\}}}
|w_{\{n\}}(t)|^2
+\dots+
2\parallel \bs O_{A_\ell}\parallel^2\sum_{\substack{n_1<...<n_s\\ n_k\in A_\ell\cup A_{\ell'},\forall k\in\{1,...,s\}}}
|w_{\{n\}}(t)|^2
\\\leq 2\parallel \bs O_{A_\ell}\parallel^2 \sum_{i=1}^{s}
\sum_{\substack{n_1<...<n_s\\n_i\in A_\ell\cup A_{\ell'} }}
|w_{\{n\}}(t)|^2 
\,,
\end{multline}
and finally
\begin{multline}
\label{eq:square value s vs UP}
|\braket{\Psi_{s}(t)|\bs O^2|\Psi_{s}(t)}-
\braket{\Uparrow|\bs O^2|\Uparrow}|
\leq
\sum_{\ell,\ell'\in\Omega_\epsilon(t)} |\braket{\Psi_{s}(t)|\bs O_{A_\ell}\bs O_{A_{\ell'}}|\Psi_{s}(t)}-
\braket{\Uparrow|\bs O_{A_\ell}\bs O_{A_{\ell'}}|\Uparrow}|
\\\leq
2\parallel\bs O_{A_1}\parallel^2
\sum_{\ell,\ell'\in\Omega_\epsilon(t)}
\sum_{i=1}^{s}
\sum_{\substack{n_1<...<n_s\\n_i\in A_\ell\cup A_{\ell'} }}
|w_{\{n\}}(t)|^2
\leq
4\parallel\bs O_{A_1}\parallel^2
\sum_{\ell,\ell'\in\Omega_\epsilon(t)}
\sum_{i=1}^{s}
\sum_{\substack{n_1<...<n_s\\n_i\in A_\ell}}
|w_{\{n\}}(t)|^2
\\\lesssim
4\parallel\bs O_{A_1}\parallel^2|A| |\Omega_\epsilon(t)|
\sum_{i=1}^{s}
\sum_{n_1<...<n_s}
|w_{\{n\}}(t)|^2
=4s\parallel\bs O_{A_1}\parallel^2|A||\Omega_\epsilon(t)|
\, .
\end{multline}
This shows that the variance of the any operator $\bs O$ on the state $\ket{\Psi_s(t)}$ coincides with the one on the state $\ket{\Uparrow}$ modulo corrections of order $\mathcal{O}(|\Omega_\epsilon(t)|)$.
Since $\ket{\Uparrow}$ is a state for which clustering of correlations holds and his variance cannot be larger than system's size, we obtain that the state $\ket{\Psi_s(t)}$ has the same properties, implying that it is not macroscopically entangled. 
Our proof can be easily generalised to any power of the operator, but higher powers are not needed to our purposes.

We now extend the results above to more general states. We  consider, first, the quantity $\braket{\Psi_{s'}|\bs O|\Psi_{s}}$, $s\neq s'$, for two different states $\ket{\Psi_s}$ and $\ket{\Psi_{s'}}$.
Without loss of generality, we can assume $s>s'$.
We start from
\begin{equation}
    |\braket{\Psi_{s'}|\bs O|\Psi_s}|\leq
    \sum_{\ell\in\Omega_\epsilon(t)} 
    \sum_{n'_1<...<n'_{s'}}\sum_{n_1<...<n_{s}} |\braket{\Uparrow|\bs\sigma^+_{n'_1}...\bs\sigma^+_{n'_{s'}} \bs O_{A_\ell} \bs\sigma^-_{n_1}...\bs\sigma^-_{n_{s}}|\Uparrow}|
    |w'_{\{n'\}}(t)| |w_{\{n\}}(t)|,
\end{equation}
and rewrite the sums regrouping them, for fixed $\ell$, according to how many of the primed indices $\{n'\}$ are outside the set $A_\ell$. This is a number $p$ that goes from 0 to $s'$. In order to get a non-zero contribution, the number of indices $\{n\}$ outside $A_\ell$ should also be $p$. The number of indices $\{n'\}\in A_\ell$ is then $s'-p\in\{0,...,s'\}$, while the number of indices $\{n\}\in A_\ell$ is $s-p\in\{s-s',...,s\}$; note that there is no contribution from  $|A_\ell|<s-p$. 
For given $p$, there are $p+1$ distinct subsets of consecutive $n$ that can be in $A_\ell$ (they are characterised by the position of the first element of the subset); to keep track of this fact, we introduce a sum over an index $j$, such that the index $n_j$ is the smallest of the indices $\{n\}\in A_\ell$. 
Such a decomposition leads to
\begin{multline}
|\braket{\Psi_{s'}|\bs O|\Psi_s}|\leq
    \sum_{\ell\in\Omega_\epsilon(t)} 
    \sum_{p=0}^{s'}
    \sum_{j=1}^{p+1}
    \sum_{\substack{n_1<...<n_{s}\\n_k\in A_{\ell}\Leftrightarrow j\leq k<j+s-p}}
    \sum_{\substack{n'_{1}<...<n'_{s'-p}\\n'_k\in A_{\ell},\forall k\in\{1,...,s'-p\}}}
    \\|\braket{\Uparrow|\bs\sigma^+_{n'_{1}}...\bs\sigma^+_{n'_{s'-p}} \bs O_{A_\ell} \bs\sigma^-_{n_j}...\bs\sigma^-_{n_{j+s-p-1}}|\Uparrow}|
    |w'_{\{n\}\setminus\{n_j,...,n_{j+s-p-1}\} \cup \{n'_1,...,n'_{s'-p}\}}(t)| |w_{\{n\}}(t)|
    \leq\\\leq
    \parallel \bs O_{A_\ell}\parallel 
    \sum_{p=0}^{s'}
    \sum_{j=1}^{p+1} \sum_{\ell\in\Omega_\epsilon(t)} 
    \sum_{\substack{n_1<...<n_{s}\\n_k\in A_{\ell}\Leftrightarrow j\leq k<j+s-p}}
    \sum_{\substack{n'_{1}<...<n'_{s'-p}\\n'_k\in A_{\ell},\forall k\in\{1,...,s'-p\}}}
    |w'_{\{n\}\setminus\{n_j,...,n_{j+s-p-1}\} \cup \{n'_1,...,n'_{s'-p}\}}(t)| |w_{\{n\}}(t)|.
\end{multline}
It is now convenient to write the sum over the indices $n,n'$ in $A_\ell$ in another way.  We introduce $P_k^K$ as the set of all the $k$-tuples of increasing numbers in $\{0,...,K-1\}$. Note that $P_k^K$ has $\binom{K}{k}$ elements. We then have
\begin{multline}
|\braket{\Psi_{s'}|\bs O|\Psi_s}|\leq
    \parallel \bs O_{A_\ell}\parallel 
    \sum_{p=0}^{s'}
    \sum_{j=1}^{p+1} \sum_{\ell\in\Omega_\epsilon(t)} 
\sum_{(k_1,...,k_{s-p})\in P_{s-p}^{|A|}}
\sum_{(k'_1,...,k'_{s'-p})\in P_{s'-p}^{|A|}}    
    \sum_{n_1<...<n_{j-1}<A_\ell<n_{j+s-p}<...<n_{s}}
    \\
    |w'_{\{n_1,...,n_{j-1},\ell+k'_1,...,\ell+k'_{s'-p},n_{j+s-p},...,n_{s}\}}(t)| |w_{\{n_1,...,n_{j-1},\ell+k_1,...,\ell+k_{s-p},n_{j+s-p},...,n_s}(t)|,
\end{multline}
which we rewrite isolating the term with $p=s'$ (the case in which all the indices $\{n'\}$ are outside $A_\ell$), which is special: 
\begin{multline}
|\braket{\Psi_{s'}|\bs O|\Psi_s}|\leq
\parallel \bs O_{A_\ell}\parallel 
    \sum_{p=0}^{s'-1}
    \sum_{j=1}^{p+1} 
\sum_{(k_1,...,k_{s-p})\in P_{s-p}^{|A|}}
\sum_{(k'_1,...,k'_{s'-p})\in P_{s'-p}^{|A|}}    
\sum_{\ell\in\Omega_\epsilon(t)} 
    \sum_{n_1<...<n_{j-1}<A_\ell<n_{j+s-p}<...<n_{s}}
    \\
    |w'_{\{n_1,...,n_{j-1},\ell+k'_1,...,\ell+k'_{s'-p},n_{j+s-p},...,n_{s}\}}(t)| |w_{\{n_1,...,n_{j-1},\ell+k_1,...,\ell+k_{s-p},n_{j+s-p},...,n_s}(t)|
    \\+
    \parallel \bs O_{A_\ell}\parallel 
    \sum_{j=1}^{s'+1} 
    \sum_{(k_1,...,k_{s-s'})\in P_{s-s'}^{|A|}}
    \sum_{\ell\in\Omega_\epsilon(t)} 
    \sum_{n_1<...<n_{j-1}<A_\ell<n_{j+s-s'}<...<n_{s}}
    \\
    |w'_{\{n_1,...,n_{j-1},n_{j+s-s'},...,n_{s}\}}(t)| |w_{\{n_1,...,n_{j-1},\ell+k_1,...,\ell+k_{s-s'},n_{j+s-s'},...,n_s}(t)|
    .
\end{multline}
Now, for fixed $p,j,\{k\},\{k'\}$, we want to  maximize the sums over $\ell,\{n\}$ independently, which gives an upper bound to the full quantity. To that aim, we introduce the following compact notation: for the terms with $p<s'$
\begin{multline}
\sum_{\ell\in\Omega_\epsilon(t)} 
    \sum_{n_1<...<n_{j-1}<A_\ell<n_{j+s-p}<...<n_{s}}
    |w'_{\{n_1,...,n_{j-1},\ell+k'_1,...,\ell+k'_{s'-p},n_{j+s-p},...,n_{s}\}}(t)|
    \times\\\times
    |w_{\{n_1,...,n_{j-1},\ell+k_1,...,\ell+k_{s-p},n_{j+s-p},...,n_s}(t)|
    \quad\equiv 
    \sum_{b\in B} W'_b W_b
    \,,
\end{multline}
and for the term $p=s'$
\begin{multline}
    \sum_{\ell\in\Omega_\epsilon(t)} 
    \sum_{n_1<...<n_{j-1}<A_\ell<n_{j+s-s'}<...<n_{s}}
    |w'_{\{n_1,...,n_{j-1},n_{j+s-s'},...,n_{s}\}}(t)| |w_{\{n_1,...,n_{j-1},\ell+k_1,...,\ell+k_{s-s'},n_{j+s-s'},...,n_s}(t)|
    \\ \equiv
    \sum_{\ell\in\Omega_\epsilon(t)}\sum_{c\in C} W'_{c}W_{c,\ell}
    \,,
\end{multline}
where $b$ and $c$ regroup all the indices that $w$ and $w'$ have in common and $W,W'$ are a synthetic expression for the coefficients that report only the indices we are interested in; note that in the first case there are as many independent $w$ as $w'$, while, for $p=s'$, $w'$ does not depend on $\ell$ anymore.
In order to maximize the expressions above, we should take into accounts the constraints  $\sum_{n'_1<...<n'_{s'}}|w'_{\{n'\}}|^2=1$ and $\sum_{n_1<...<n_{s}}|w_{\{n\}}|^2=1$. In our search for the maximum, we can use the stronger conditions $\sum_{b\in B}W_b^2=1$ and $\sum_{b\in B}(W'_b)^2=1$ in the first case and $\sum_{c\in C}(W_c')^2=1$ and $\sum_{c\in C,\ell\in \Omega_\epsilon(t)}W_{c,\ell}^2=1$ in the second case, which amount to require that the coefficients that are not involved in the sum are zero (given the original constraint, this maximises the sum because all the terms are positive).
Using the method of Lagrange multipliers, one can show that, under these constraints, $\sum_{b\in B} W'_b W_b$ is maximised by $W_b=W'_b$ and equals 1, while  $\sum_{\ell\in\Omega_\epsilon(t)}\sum_{c\in C} W'_{c}W_{c,\ell}$ has degenerate points of maximum in $W_{c,\ell} = |\Omega_\epsilon(t)|^{-1/2} W'_c$ and gives $|\Omega_\epsilon(t)|^{1/2}$.
In the end 
\begin{multline}
\label{eq:overlap_O}
|\braket{\Psi_{s'}|\bs O|\Psi_s}|\leq
    \parallel \bs O_{A_\ell}\parallel 
    \sum_{j=1}^{s'+1} 
    \sum_{(k_1,...,k_{s-s'})\in P_{s-s'}^{|A|}} |\Omega_\epsilon(t)|^{1/2}
        + \mathcal{O}(|\Omega_\epsilon(t)|^0)
        =\\=
         \parallel \bs O_{A_\ell}\parallel
    (s'+1) \binom{|A|}{s-s'} |\Omega_\epsilon(t)|^{1/2} 
    + \mathcal{O}(|\Omega_\epsilon(t)|^0)
    \,.
\end{multline}
This result will be used below.

We finally consider a finite sum of states obtained by flipping different numbers of spins: $\ket{\Psi^{(S)}} = \sum_{s=1}^S a_{s}\ket{\Psi_{s}}$, where $a_j\in\mathbb{R}, \forall j$ (any potential phase can be absorbed in the states), and $\sum_{s=1}^S a_j^2=1$.
From Eq.~[\ref{eq:overlap_O}] it readily follows
\begin{equation}
    |\braket{\Psi_{S+1}|\bs O|\Psi^{(S)}}|
    \leq
    \sum_{s=1}^S |\braket{\Psi_{S+1}|\bs O|\Psi_{s}}|
    \lesssim
    \mathcal{O}(|\Omega_\epsilon(t)|^{1/2})
\,.
\end{equation}
Let us now consider the state $\ket{\Psi^{(2)}} = a_1\ket{\Psi_1} + a_2 \ket{\Psi_{2}}$, with $a_1,a_2\in\mathbb{R}$ and $a_1^2+a_2^2=1$. First of all, note that such a state is macroscopically equivalent to $\ket{\Uparrow}$. Indeed, using $\braket{\Psi_{2}|\bs O|\Psi_{2}}=\braket{\Psi_{1}|\bs O|\Psi_{1}}+\mathcal{O}(|\Omega_\epsilon(t)|^0)$ (a corollary of Eq.~[\ref{eq:expectation value s vs UP}]), we have
\begin{multline}
    \braket{\Psi^{(2)}|\bs O|\Psi^{(2)}} - \braket{\Uparrow|\bs O|\Uparrow}
    = a_1^2\braket{\Psi_{1}|\bs O|\Psi_{1}} + a_2^2\braket{\Psi_{2}|\bs O|\Psi_{2}} + 2 a_1a_2 Re(\braket{\Psi_{1}|\bs O|\Psi_{2}})- \braket{\Uparrow|\bs O|\Uparrow}
    \\= 2 a_1a_2 Re(\braket{\Psi_{1}|\bs O|\Psi_{2}}) + \mathcal{O}(|\Omega_\epsilon(t)|^0)
     \lesssim \mathcal{O}(|\Omega_\epsilon(t)|^{1/2})\,,
\end{multline}
which is sub-extensive. Similarly, we can conclude that $\braket{\Psi^{(S)}|\bs O|\Psi^{(S)}} - \braket{\Uparrow|\bs O|\Uparrow}\lesssim \mathcal{O}(|\Omega_\epsilon(t)|^{1/2})$, meaning that $\ket{\Psi^{(S)}}$ is macroscopically equivalent to $\ket{\Uparrow}$ for any finite $S$. 
Let us now look at the variance of $\bs O$ in $\ket{\Psi^{(2)}}$:
\begin{multline}
    \braket{\Psi^{(2)}(t)|\bs O^2|\Psi^{(2)}(t)} - \braket{\Psi^{(2)}(t)|\bs O|\Psi^{(2)}(t)}^2
     = a_1^2 \braket{\Psi_{1}|\bs O^2|\Psi_{1}} + a_2^2 \braket{\Psi_{2}|\bs O^2|\Psi_{2}} + 2a_1 a_2 Re\left(\braket{\Psi_{1}|\bs O^2 |\Psi_{2}}\right)
    +\\-\left(a_1^2 \braket{\Psi_{1}|\bs O|\Psi_{1}} + a_2^2 \braket{\Psi_{2}(t)|\bs O|\Psi_{2}(t)} + 2a_1a_2 Re\left(\braket{\Psi_{1}|\bs O |\Psi_{2}(t)}\right)\right)^2
    =\\ = \braket{\Psi_{1}|\bs O^2|\Psi_{1}} + 2a_1a_2 Re\left(\braket{\Psi_{1}|\bs O^2 |\Psi_{2}(t)}\right)+ \mathcal{O}(|\Omega_\epsilon(t)|)
    +\\-\left(\braket{\Psi_{1}|\bs O|\Psi_{1}} + 2a_1a_2 Re\left(\braket{\Psi_{1}|\bs O |\Psi_{2}(t)}\right) + \mathcal{O}(|\Omega_\epsilon(t)|^0)\right)^2 
    \,,
\end{multline}
where we used $\braket{\Psi_{2}|\bs O|\Psi_{2}}=\braket{\Psi_{1}|\bs O|\Psi_{1}}+\mathcal{O}(|\Omega_\epsilon(t)|^0)$ and $\braket{\Psi_{2}|\bs O^2|\Psi_{2}}=\braket{\Psi_{1}|\bs O^2|\Psi_{1}}+\mathcal{O}(|\Omega_\epsilon(t)|)$ (corollaries of Eq.~[\ref{eq:expectation value s vs UP}] and Eq.~[\ref{eq:square value s vs UP}]).
Using that the variance of any state $\ket{\Psi_s}$ can not grow faster than $|\Omega_\epsilon(t)|$, we get
\begin{multline}
    \braket{\Psi^{(2)}(t)|\bs O^2|\Psi^{(2)}(t)} - \braket{\Psi^{(2)}(t)|\bs O|\Psi^{(2)}(t)}^2
    \\=2a_1a_2 \left[ Re\left(\braket{\Psi_{1}|\bs O^2 |\Psi_{2}(t)}\right) -\braket{\Psi_{1}|\bs O|\Psi_{1}}Re\left(\braket{\Psi_{1}|\bs O |\Psi_{2}(t)}\right)\right] + \mathcal{O}(|\Omega_\epsilon(t)|)
    \,.
\end{multline}
If the leading contribution to the variance were larger than $\mathcal{O}(|\Omega_\epsilon(t)|)$,  by changing the sign of $a_1$ we could make the variance negative; since the variance is by definition positive, by contradiction, we have that the leading order cannot be larger than $\mathcal{O}(|\Omega_\epsilon(t)|)$.
We can now assume that the variance in the state $\ket{\Psi^{(S-1)}}$ is of order $\mathcal{O}(|\Omega_\epsilon(t)|)$ and show that this implies that also the variance in $\ket{\Psi^{(S)}}$ is.
The proof is essentially the same as above with just a small modification, namely, $\braket{\Psi^{(S-1)}|\bs O|\Psi^{(S-1)}}-\braket{\Uparrow|\bs O|\Uparrow}$ is $\mathcal{O}(|\Omega_\epsilon(t)|^{1/2})$ rather than $\mathcal{O}(|\Omega_\epsilon(t)|^{1/2})$. This gives an extra term in the potential leading order, which however does not affect the final result: calling $\ket{\Psi^{(S)}}=a_1\ket{{\Psi_S}} + a_2\ket{\Psi^{(S-1)}}$, we have
\begin{multline}
    \braket{\Psi^{(S)}(t)|\bs O^2|\Psi^{(S)}(t)} - \braket{\Psi^{(S)}(t)|\bs O|\Psi^{(S)}(t)}^2
    \\ = a_1^2 \braket{\Psi_S|\bs O^2|\Psi_S} + a_2^2 \braket{\Psi^{(S-1)}|\bs O^2|\Psi^{(S-1)}} + 2a_1 a_2 Re\left(\braket{\Psi_S|\bs O^2 |\Psi^{(S-1)}}\right)
    \\-\left(a_1^2 \braket{\Psi_S|\bs O|\Psi_S} + a_2^2 \braket{\Psi^{(S-1)}(t)|\bs O|\Psi^{(S-1)}(t)} + 2a_1a_2 Re\left(\braket{\Psi_S|\bs O |\Psi^{(S-1)}(t)}\right)\right)^2
    \\ = \braket{\Psi_S|\bs O^2|\Psi_S} + 2a_1a_2 Re\left(\braket{\Psi_S|\bs O^2 |\Psi^{(S-1)}(t)}\right)+ \mathcal{O}(|\Omega_\epsilon(t)|)
    \\-\left(\braket{\Psi_S|\bs O|\Psi_S} + 2a_1a_2 Re\left(\braket{\Psi_S|\bs O |\Psi^{(S-1)}(t)}\right) + \mathcal{O}(a_1^0|\Omega_\epsilon(t)|^{1/2})\right)^2 
    \\\lesssim 2a_1a_2 \left[ Re\left(\braket{\Psi_S|\bs O^2 |\Psi^{(S-1)}(t)}\right) -\braket{\Psi_S|\bs O|\Psi_S}Re\left(\braket{\Psi_S|\bs O |\Psi^{(S-1)}(t)} +  \mathcal{O}(a_1^0|\Omega_\epsilon(t)|^{1/2}\right)\right] + \mathcal{O}(|\Omega_\epsilon(t)|)
    \,.
\end{multline}
Since $a_1$ appears linearly as in the case $S=2$, we can use the same trick and we have by contradiction that the variance in $\ket{\Psi^{(S)}}$ cannot grow faster than the system's size.

In conclusion, we have proved that the state obtained by any local perturbation of $\ket{\Uparrow}$ is macroscopically equivalent to $\ket{\Uparrow}$ and the variance of any operator in this state does not grow faster than system size, where the system size is defined by $\Omega_\epsilon(t)$.

\end{document}